\documentclass[fleqn,usenatbib]{mnras}

\usepackage{newtxtext,newtxmath}
\usepackage[ampersand]{easylist}
\usepackage{ulem}

\usepackage[T1]{fontenc}
\usepackage{ae,aecompl}

\usepackage{graphicx}	
\usepackage{amsmath}	
\usepackage{amssymb}	
\usepackage[colorinlistoftodos]{todonotes}

\newcommand{\Msun}{\ensuremath{{\rm M}_{\odot}}}

\newcommand{\Ks}{$K_s$}
\newcommand{\K}{$K_s$}
\newcommand{\gzK}{$gzK_s$}
\newcommand{\PEgzK}{$P\!E$-$gzK_s$}
\newcommand{\SFgzK}{$SF\!$-$gzK_s$}
\newcommand{\zs}{$z$$\sim$}

\newcommand{\Mstars}{$M_{\star}$}
\newcommand{\sqdeg}{deg$^2$}

\newcommand{\MXXL}{{\it {MXXL}}}
\newcommand{\peryr}{yr$^{-1}$}

\title[LARgE -- II. Dark Halos of $z \sim 1.6$ UMPEGs]{LARgE Survey -- II. The Dark Matter Halos and the Progenitors and Descendants of Ultra-Massive Passive Galaxies at Cosmic Noon}

\author[G.K. Cheema et al.]{\hspace{-2.1mm}
Gurpreet Kaur Cheema$^{1}$\thanks{E-mail: {\tt gcheema@ap.smu.ca}},
Marcin Sawicki$^{1,2}$\thanks{E-mail: {\tt marcin.sawicki@smu.ca}}\thanks{Canada Research Chair},
Liz Arcila-Osejo$^{1}$,
Anneya Golob$^{1}$,
\newauthor
Thibaud Moutard$^{1}$,
St\'ephane Arnouts$^{3}$,
Jean Coupon${^4}$
\\$^{1}$Institute for Computational Astrophysics and Department of Astronomy \& Physics, Saint Mary's University, 923 Robie Street, Halifax, NS B3H 3C3, Canada
\\$^{2}$NRC Herzberg Astronomy and Astrophysics, 5071 West Saanich Road, Victoria, BC V9E 2E7, Canada 
\\$^{3}$Laboratoire d'Astrophysique de Marseille, 38 rue Frederic Joliot Curie, Universit\'e Aix-Marseille, Marseille, F-13388, France
\\$^{4}$Astronomy Department, University of Geneva, Chemin d'Ecogia 16, CH-1290 Versoix, Switzerland
}

\date{Accepted for publication in MNRAS}

\pubyear{2020}

\begin{document}
\label{firstpage}
\pagerange{\pageref{firstpage}--\pageref{lastpage}}
\maketitle
\begin{abstract}
We use a 27.6~\sqdeg\ survey to measure the clustering of \gzK-selected quiescent galaxies at $z\sim1.6$, focusing on ultra-massive quiescent galaxies. We find that $z\sim1.6$ Ultra-Massive Passively Evolving Galaxies (UMPEGs), which have $K_s(AB)<19.75$ (stellar masses of \Mstars $\ga 10^{11.4}M_{\odot}$ and mean  $<$\Mstars$>$ = $10^{11.5}M_{\odot}$), cluster more strongly than any other known galaxy population at high redshift. Comparing their correlation length, $r_0 = 29.77 \pm 2.75$
 $ h^{-1}$Mpc, with the clustering of dark matter halos in the {\it Millennium XXL} N-body simulation suggests that these $z\sim1.6$ UMPEGs reside in dark matter halos of  mass $M_{h}\sim10^{14.1}h^{-1}M_{\odot}$.  Such very massive $z\sim1.6$ halos are associated with the ancestors of $z\sim0$ massive galaxy clusters such as the Virgo and Coma clusters. Given their extreme stellar masses and lack of companions with comparable mass, we surmise that these UMPEGs could be the already-quenched central massive galaxies of their (proto)clusters. We conclude that with only a modest amount of further growth in their stellar mass, $z\sim1.6$ UMPEGs could be the progenitors of some of the massive central galaxies of present-day massive galaxy clusters observed to be already very massive and quiescent near the peak epoch of the cosmic star formation.
\end{abstract}







\section{Introduction\label{sec:Introduction}}

According to the standard $\Lambda$CDM cosmological model, the matter content of the Universe is dominated by cold dark matter (CDM). The growth of the first gravitational instabilities that arose from primordial quantum fluctuations being rapidly and exponentially inflated to much larger sizes leads to the gravitational collapse of over-dense regions of dark matter (DM) into the first DM halos. Subsequently, baryonic matter falls into the  gravitational wells of the DM halos and turns into luminous galaxies via the cooling and condensation of baryons \citep*{doi:10.1093/mnras/183.3.341}. 
Following their formation, galaxies grow further through the merging of the dark matter halos and the associated baryonic material, progressively assembling into more massive systems.  The evolution of observable galaxies within their host DM halos involves various internal  and external processes such as gas cooling, hydrodynamical effects, star formation, mergers, and feedback mechanisms, all of which are linked to the properties of  the host DM halos (\citealt{0004-637X-717-1-379,doi:10.1093/mnras/stv1438}). Since the properties of galaxies are directly coupled to the properties of the DM halos in which they reside, they will  also change over cosmic time as their halos grow. Consequently, galaxies at high redshift can be expected to be different from those at the present epoch.

The most massive halos can be expected to host some of the most massive galaxies and so,  in \cite{LARGE1},  we assembled a sample of such massive, quiescent galaxies at \zs1.6. Here we define Ultra Massive Passively Evolving Galaxies (UMPEGs) to be extreme galaxies at $z\sim1.6$, with stellar masses \Mstars$>10^{11.4}M_{\odot}$ (some as massive as \Mstars$\sim10^{11.8}$\Msun)\footnote{Our \Mstars$>10^{11.4}M_{\odot}$ UMPEG mass cut is marginally lower than the  \Mstars$>10^{11.5}M_{\odot}$  used in \cite{LARGE1} so as to allow us a larger sample that's required for clustering analysis.};  critically, our UMPEGs appear to be very rare systems that are no longer forming new stars. The combination of their very high stellar masses and their low (or non-existent) star formation rates makes UMPEGs extremely rare at this redshift. Since the Universe was only $\sim$4 Gyr old at $z\sim 1.6$, their massive stellar populations must have assembled very early and rapidly. Assuming that these galaxies were on or above the star-forming main sequence \citep[e.g.,][]{Whitaker2012} just before becoming quenched, they must have exhibited star formation rates of SFR$\sim$200--1000~M$_{\odot}$\peryr.  Moreover, our  UMPEGs have very high stellar masses and have no companions of equal or higher mass. They also  have virtually no satellites with stellar masses down to $\sim$1/3 times the mass of the UMPEG (M.~Sawicki et al., submitted to MNRAS). Consequently, they can be expected to be the most massive, central galaxies of their dark matter halos.

While massive and  bright (\Ks$\sim$19.5 AB), UMPEGs are exceedingly rare and populate the very massive, exponential tail end of the $z\sim1.6$ galaxy stellar mass function (SMF).  The number density of UMPEGs is $\sim$$10^{-6}$ Mpc$^{-3}$ per dex in stellar mass  (see \citealt{LARGE1}), which is two orders of magnutide lower than that of the $z\sim1.6$ typical $M_\star^*$$\sim$$10^{10.7}M_{\odot}$ galaxies that are normally considered to be ``massive galaxies'' at these redshifts.  Because they are extreme systems, UMPEGs can  be used to test the  extremes of the hierarchical models of galaxy formation and  evolution.  For example, \citet{LARGE1} showed that the \zs1.6 quiescent galaxy SMF follows very closely the \citet{Schechter1976} functional form over \Mstars$ \sim 10^{10.2} -10^{11.7}$\Msun\ -- a factor of 30 in mass -- in excellent agreement with predictions of ``mass quenching'' models \citep[e.g.,][]{Peng2010}.

The extreme nature of UMPEGs raises the question: what environments do they reside in?  One way to address this question is by examining their clustering properties. Previous studies have found that galaxy properties such as stellar mass, luminosity, morphology and star formation rate are correlated with host DM halo mass (e.g., \citealt{Li2006,Zehavi2011}), highlighting the fact that the DM halo environment plays an important role in shaping the  properties of galaxies and thus galaxy evolution in general.  The DM halo mass can also be helpful in tracing the mass assembly history of these extreme galaxies because dark matter halo growth is well understood from N-body simulations \citep[e.g.,][]{2005Natur.435..629S,doi:10.1111/j.1365-2966.2009.15191.x} and is independent of the complicated and poorly-understood baryonic processes inside the halos.

If UMPEGs -- which have very high stellar masses -- are associated with massive DM halos, then we would expect that the clustering of  UMPEGs, reflecting the underlying clustering of their halos, will be stronger than the clustering of ``normal'' massive quiescent galaxies at the same epoch. In this paper we aim to test this scenario and constrain the masses of the UMPEG halos by quantitatively comparing their clustering with predictions of dark matter clustering from N-body simulations \citep[e.g.,][]{2005Natur.435..629S,doi:10.1111/j.1365-2966.2009.15191.x}.   Several alternative methods, such as halo occupation distribution modelling \citep[e.g.,][]{0004-637X-738-1-45,2003MNRAS.339.1057Y,0004-637X-575-2-587}, the stellar mass-halo mass relation, halo (or sub-halo) abundance matching \citep[e.g.,][]{0004-637X-520-2-437,0004-637X-647-1-201,2010ApJ...710..903M}, and weak gravitational lensing \citep[e.g.,][]{1996ApJ...466..623B,0004-637X-606-1-67} can also statistically connect a population of galaxies to their host dark matter halos.  However, the dataset we use to identify our UMPEGS, while covering a wide area that is necessary to find these rare \zs1.6 objects in sufficient numbers, is not deep enough for either lensing or halo occupation distribution analysis at these redshifts. Meanwhile, abundance matching, which requires a complete catalog ordered by galaxy stellar mass, cannot be used reliably given that our UMPEGs are quiescent while not all galaxies with UMPEG-like masses are so. Consequently, given the limitations of our data, the auto-correlation function presents itself as the best way to constrain UMPEG halo masses at this point. 

Clustering is a powerful way to investigate the halo masses of UMPEG halos since the amplitude of clustering on large scales can provide a measure of the mass of the host DM halos \citep{doi:10.1093/mnras/282.2.347,doi:10.1046/j.1365-8711.1999.02692.x}.  In the $\Lambda CDM$ model, the clustering of halos is well understood \citep*{doi:10.1046/j.1365-8711.2002.05723.x} and the clustering amplitude is a monotonically increasing function of the halo mass. This arises because halos located in large-scale positive density perturbations form first and this early boost accelerates the collapse of halos, favours subsequent merging, and leads to an overabundance of  massive haloes in dense large-scale environments \citep{1984ApJ...284L...9K, Bond1991}.  For these reasons galaxy clustering studies are a popular and useful tool that has been used by many authors at both low and high redshifts \citep[e.g.,][]{
LeFevre1996, 
Shepherd2001, 
0004-637X-597-1-225, 
ref_brown, 
doi:10.1046/j.1365-8711.2003.06861.x,
0004-637X-609-2-525,0004-637X-630-1-1,
McCracken2008, 
0004-637X-737-2-92,
doi:10.1093/mnras/stv1927, 
Lin2016, 
Cameron2019}.  In this paper we extend this approach to our UMPEGs, which are high-redshift galaxies that are quiescent and much more massive than those previously studied at $z>1$. 

This paper is structured as follows. In Section~\ref{sec:Data} we briefly describe our data as well as the method we used in \citet{LARGE1} to select our passive high-redshift galaxies. In Section~\ref{sec:ACF}, we describe the technique we use to measure the angular correlation function for these galaxies.  In Section~\ref{sec:Spatial-Correlation-Function} we convert the angular correlation function into the spatial one using the estimated photometric redshift distribution of the passive galaxies. In \S~\ref{sec:haloMasses}  we compare the clustering results of our UMPEGs with the clustering measurements of the dark matter halos from the $Millennium$ $XXL$ simulation and thereby constrain their host DM halo masses. A discussion and interpretation of our measurements is presented in \S~\ref{sec:discussion}, and we summarize the main conclusions in \S~\ref{sec:summary}.

Throughout this work we assume the flat $\Lambda$ cosmology with $\Omega_{m}=0.3$,  $\Omega_{\Lambda}=0.7$. The Hubble constant is 
$H_{0} = 70$ km s$^{-1}$Mpc$^{-1}$ so that $h$ = $H_0 /(100$ km s$^{-1}$Mpc$^{-1})$ =  0.7, and the normalization of the matter power spectrum is $\sigma_{8}=0.8$. We use the AB magnitude system \citep{1974ApJS...27...21O} and  stellar masses of galaxies assume the \citet{1538-3873-115-809-763} stellar initial mass function (IMF).

\begin{figure*}
\includegraphics[width=18cm]{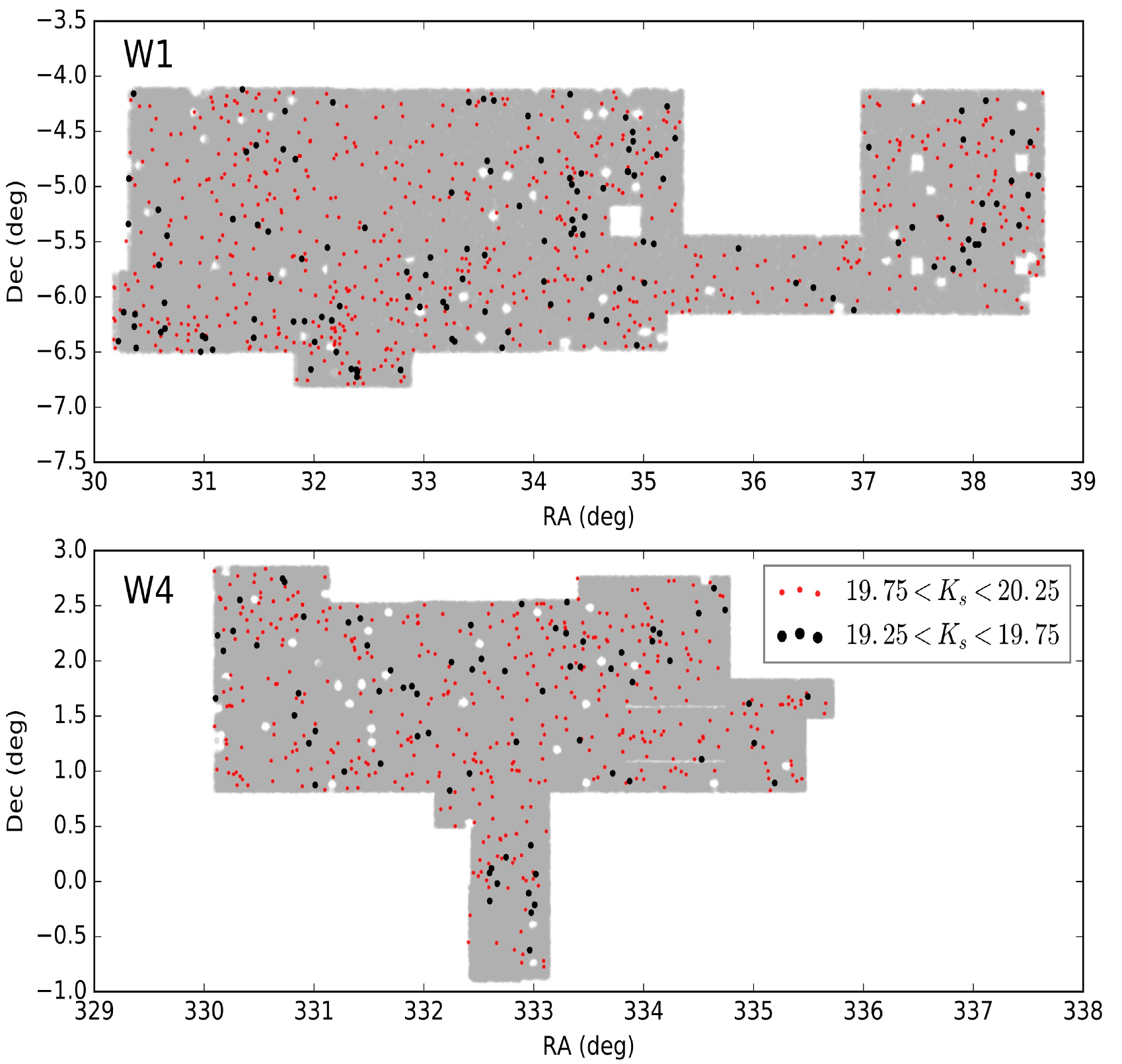}
\protect\caption[Distribution of the PE-$gzK_s$ in the Wide fields]{\label{fig:PE_wide}
Distribution of bright \PEgzK\ galaxies in the Wide fields. The gray area shows the geometry of the fields; white spaces are areas that have no data or that are masked due to bright stars or artifacts. Positions of UMPEGs are shown with black points and those of fainter \PEgzK\ galaxies with red.}
\end{figure*}

\begin{figure*}
\includegraphics[width=14cm,height=14cm,keepaspectratio]{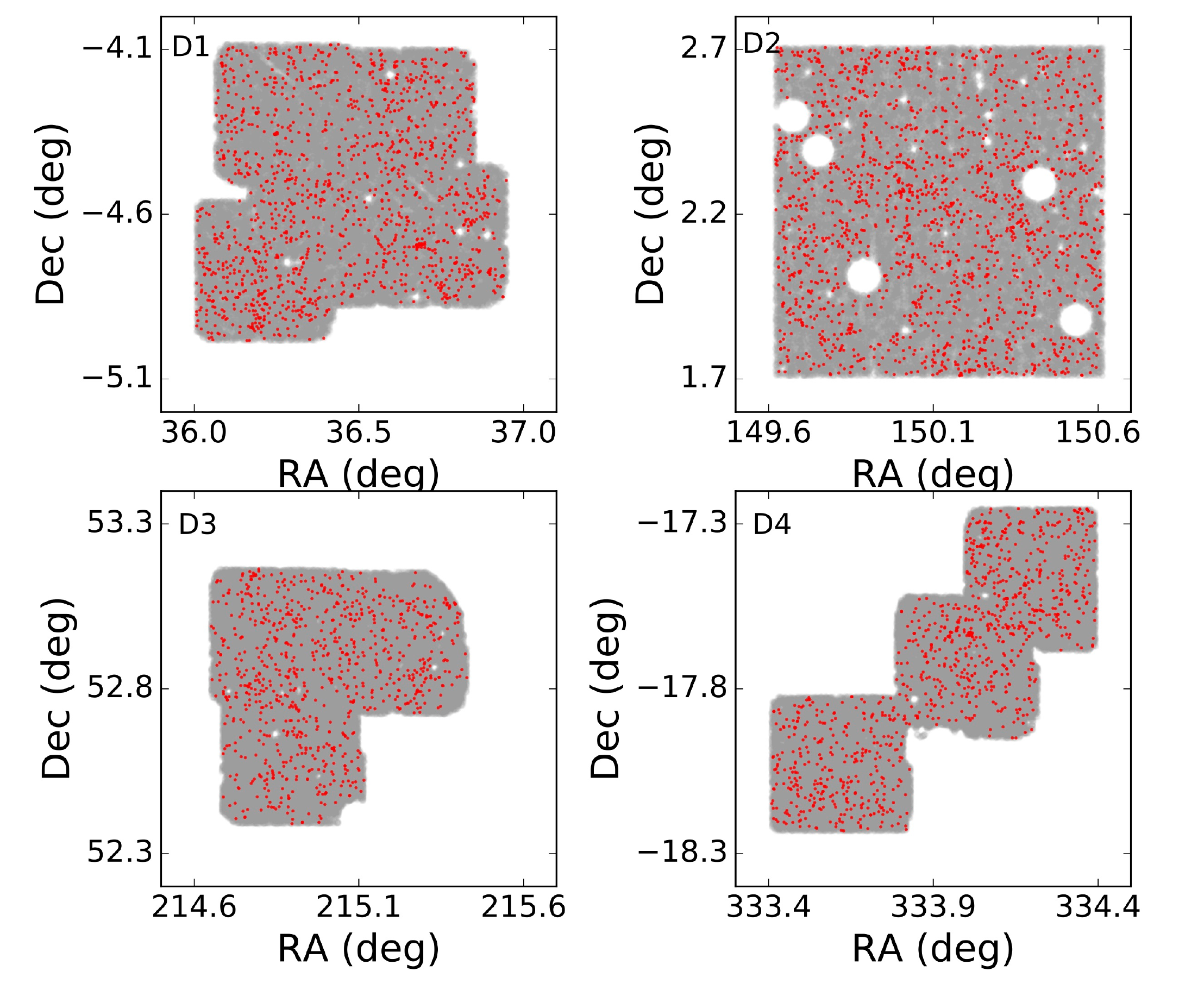}
\protect\caption[Distribution of the PE-$gzK_s$ in the Deep fields]
{\label{fig:deep-survey} Distribution of the PE-$gzK_s$ galaxies in the Deep fields. The gray area shows the geometry of the fields; white spaces are areas that have no data or that are masked due to bright stars or artifacts. Note that these four fields are much smaller than the Wide fields shown in Fig.~\ref{fig:PE_wide}, with each panel here showing just $1.2\deg \times 1.2\deg$.}
\end{figure*}

\section{Data and sample selection}\label{sec:Data}

The data and high-redshift galaxy selection are described in detail in \cite{LARGE1}, and thus here we provide only a brief overview.  

\subsection{Data}

As described in more detail in Sec.~\ref{sec:gzKselection}, we select high-$z$ passive galaxies using an adaptation of the \cite{0004-637X-617-2-746} $BzK$ technique developed by \cite{ArcilaOsejo2013}.  This technique requires near-infrared (NIR) as well as optical photometry, specificall $g$, $z$, and $K_s$ fluxes, in the Deep fields of our survey (see below) also supplemented with $H$-band measurements. 

For the optical data ($g$, $z$) we use the images from the T0006 release of the CFHT Legacy Survey (CFHTLS, \citealt{CFHTLS-T0006}) -- specifically the $g$ and $z$ images of all four of its Deep fields (D1, D2, D3, and D4), and two of its Wide fields (W1 and W4).  For the NIR data,  in the Wide fields we turn to the $K_s$ images from the Visible Multi-Object Spectrograph (VIMOS) Public Extragalactic $K_{s}$ Survey Multi-Lambda Survey (VIPERS-MLS; \citealt{refIda}), while in the Deep fields we use the \Ks\ and $H$-band data from the T0002 release of the WIRCam Deep Survey (WIRDS;  \citealt{refIdb}).

The extent of the NIR images dictates our areal coverage since their footprint is smaller than that of the CFHTLS optical data.  After masking areas around bright stars, low-SNR regions, and other artifacts, our usable data cover 25.09 deg$^2$ in the two CFHTLS Wide fields and 2.51 deg$^2$ in the Deep fields, giving a total of 27.6 deg$^2$.  In the Wide fields we reach 90\% detection completeness at \Ks=20.5 AB;  in the Deep D1, D3, and D4 fields we reach 50\% detection completeness at \Ks=23.5 AB, while in D2 (the COSMOS field) we reach \Ks=23.0 AB.  

We performed source detection and photometry using SExtractor \citep{refId10}. As the Spectral Energy Distributions (SEDs) of passive galaxies are expected to be dominated by optically faint, red, long-lived stars, object detection was done in the $K_{s}$-band. Matched-aperture photometry was then done in all the bands using SExtractor's dual image mode.

\subsection{Selection of \zs1.6 Passive \gzK\ Galaxies}\label{sec:gzKselection}

We  use an adaptation of the \cite{0004-637X-617-2-746} $BzK$ technique to select high-redshift galaxies, as described in \cite{ArcilaOsejo2013} and \cite{LARGE1}. In brief, passive galaxies at $z\sim 1.5-2.5$ are red in both $g - z$ and $z - K_s$. Consequently, they are readily distinguished from other types of objects:  from star-forming galaxies at similar redshifts, which -- if dusty -- can also be red in $z - K_s$, but are blue in $g - z$;  from lower-redshift galaxies, which are bluer in $z - K_s$;  and from Galactic stars, which are bluer still. In the Deep fields, where even the very deep CFHTLS $g$ observations are often too shallow to yield a precise $g-z$ color, we additionally use the $H - K_s$ colour to break the degeneracy between high-redshift passive and star-forming galaxies (see \citealt{ArcilaOsejo2013} for details). We refer to the quiescent galaxies selected using this technique as \PEgzK\ galaxies, and to their star-forming counterparts as \SFgzK\ galaxies.

After applying the selection procedure described above to our photometric catalog, we are left with a sample of 1312 \PEgzK\ galaxies with 19.25$<$\Ks$<$20.25 in the Wide fields and 5005 \PEgzK\ galaxies with 20$<$\Ks$<$23  in the Deep fields.  In the Wide fields, 203 of these \PEgzK\ galaxies have \Ks$<$19.75, corresponding to stellar masses \Mstars$\ga$$10^{11.4}$\Msun (with mean stellar mass of  $<$\Mstars$>$ = $10^{11.5}M_{\odot}$), and these objects constitute our UMPEG sample for the present analysis. In this paper we do not study UPMEGs in the Deep fields due to these fields' small areas, although we make use of the Deep fields to measure the clustering of lower-mass quiescent \PEgzK\ galaxies.  For full details of the color-color selection, catalog creation, and the quiescent galaxy stellar mass function (SMF), see \cite{LARGE1}.

Figures~\ref{fig:PE_wide} and \ref{fig:deep-survey} show the positions of the \PEgzK\ galaxies in the Wide and Deep fields, respectively. In Fig.~\ref{fig:PE_wide} (the Wide fields) we show all \PEgzK\ objects  with $19.25<K_{s}<20.25$, with the UMPEGs ($K_{s}<19.75$) marked in black. In Fig.~\ref{fig:deep-survey} (the Deep fields) we show all \PEgzK\ objects  with $20<K_{s}<23$.  As Figs.~\ref{fig:PE_wide} and \ref{fig:deep-survey} show, \PEgzK\ galaxies appear to be clustered, and this clustering is particularly strong for the very bright \PEgzK\ galaxies (i.e., UMPEGs) in the Wide fields shown in Fig.~\ref{fig:PE_wide}.

\section{The Angular Correlation Function}\label{sec:ACF}

Once we have the positions of galaxies in our survey, the next step is to quantify their clustering properties. This is commonly done by means of the galaxy-galaxy correlation function, first suggested by \citet*{1969PASJ...21..221T} and subsequently developed for the statistical characterization of galaxy clustering \citep[e.g.,][]{PeeblesBook, ref_maddox,ref_york}.  Proceeding in the standard way we first measure the two-dimensional correlation function (this Section) and then infer the three-dimensional correlation function by statistically de-projecting the two-dimensional one in Sec.~\ref{sec:Spatial-Correlation-Function}.  Users familiar with two-point correlation function measurements may find it expedient to skip Sec.~\ref{sec:LSestimator} (which describes the details of the \citealt{1993ApJ...412...64L} clustering estimator, jackknife uncertainty measurements, and the integral constraint) and proceed directly to Sec.~\ref{sec:2Dresults} for our 2-D clustering results. Similarly, readers familiar with the Limber inversion may want to skip Sec.~\ref{sec: limber inversion}, which describes the principle of that technique, and proceed to Sec.~\ref{sec:redshift distributions}.

\subsection{Procedure}\label{sec:LSestimator}

The angular two-point correlation function is defined as the joint probability $dP(\theta)$ of finding a pair of objects in the solid angles $d\Omega_{1}$ and $d\Omega_{2}$ separated by an angle $\theta$, as compared with an unclustered random distribution and is written as
\begin{equation}
dP(\theta)=n[1+\omega(\theta)]d\Omega_{1}d\Omega_{2},\label{eq:angular correlation}
\end{equation}
where $n$ is the average surface density of galaxies and $\omega(\theta)$ is the two-point correlation function. Thus, $\omega(\theta)$ describes, as a function of angular separation $\theta$,  the excess clustering of galaxies compared to a random distribution. A positive $\omega(\theta)$ indicates that objects are clumped relative to a random distribution.

\subsubsection{Correlation function estimator}

Operationally, we measure the angular clustering using the \citeauthor{1993ApJ...412...64L} (1993, hereafter LS) estimator. 
Although computationally expensive compared to other methods \citep{1974ApJS...28...19P, 1983ApJ...267..465D, 1982MNRAS.201..867H,1993ApJ...417...19H, 1993ApJ...412...64L}, the LS estimator has several advantages: it has superior shot-noise behaviour  \citep{1538-4357-494-1-L41}, low sensitivity to the size of the random catalog, handles survey-edge corrections well \citep{Kerscher2000}, and has minimal variance for a random distribution \citep{Labatie201285}.

The LS estimator is given by
\begin{equation}
\label{eq:LS}
\omega(\theta)=\frac{DD(\theta) - 2 DR(\theta)+RR(\theta)}{RR(\theta)},
\end{equation}
where $DD(\theta)$ is the number of unique galaxy-galaxy pairs with angular separations between $\theta-\Delta\theta/2$ and $\theta+\Delta\theta/2$; $DR(\theta)$ is the number of pairs with the same angular separations between the observed galaxy catalog and the catalog of randomly positioned points in the same survey area; and $RR(\theta)$ refers to the number of random-random pairs with the same angular separations. While $DD$ measures the clustering of galaxies within the survey, $RR$ and $DR$ account for the geometry and (position-dependent) depth of the survey.

\subsubsection{Correlation function measurement and its uncertainty \label{sub:Error-Estimation_one_halo}}

We generated the random catalog needed for the $DR$ and $RR$ measurements by randomly sampling positions within the survey area, subject to bright star and artifact masks described in Section~\ref{sec:gzKselection}.  Because our object catalogs are well above the detection limits, we do not need to account for position-dependent survey depth. To suppress Poisson noise in the $DR$ and $RR$ terms in Eq.~\ref{eq:LS}, our random catalog contains $\sim$100 times more positions than the galaxy catalog (i.e., $N_{R}/N_{D}\sim100$).   So as not to give undue weight to the (oversampled) random points, we apply weighing to the terms in Eq.~\ref{eq:LS}: DR is multiplied by $(N_{D}-1)/N_{R}$ and RR by $[N_{D}(N_{D}-1)]/[N_{R}(N_{R}-1)]$ \citep{Adelberger2005}. 

The number of pairs is large, particularly for the $DR$ and $RR$ terms. For $N_D$ galaxies, there are $\frac{1}{2}N_D(N_D-1)$ $\approx$ $\frac{1}{2}N_D^{1/2}$  data-data pairs in the survey, and 
many more data-random and random-random pairs given that $N_R\sim 100 N_D$. Counting the number of pairs in each angular separation bin is thus a challenge to computational power and memory. We solve this problem by organizing our pairs catalogs using $kd$ trees \citep{Friedman:1977:AFB:355744.355745} to pre-sort our pairs in a way that allows quick identification of pairs with separations in the desired $\theta\pm\Delta\theta/2$ bin.

Uncertainties in clustering measurement can be estimated using simple error propagation that assumes Gaussian statistics in DD, DR, and RR pair counts in each bin \citep{1993ApJ...412...64L}.  However, a more accurate approach is to estimate uncertainties using a data-resampling technique, such as jackknife resampling, because the fact that uncertainties in DD, DR, and RR are not independent can lead to biased results in the classical error-propagation approach.  

We thus use jacknnife resampling. To estimate jackknife uncertainties, the data in each field are divided into grids of $N$ sub-areas ($N$=12 or 16 in the Wide fields) and the uncertainty $\sigma$, is estimated from the scatter in $N$ measurements, each of which excludes the $i$th sub-area.  This can be written as
\begin{equation}\label{eq:jackknifeError}
\sigma^2(\theta) = \sum^N_{i=1}\frac{DR_{i}(\theta)}{DR(\theta)}\left[\omega_{i}(\theta) - \omega(\theta)\right]
\end{equation}
\citep{Nikoloudakis2013}, where $w(\theta)$ is measured using the whole sample in the given field, $w_{i}(\theta)$ is measured using the whole sample in the field but excluding the $i$th subarea, and $DR_{i}(\theta)/DR(\theta)$, which is slightly smaller than unity, accounts for the fact that each $i$th measurement excludes one of the $N$ subareas. 

Our correlation function measurement in each angular bin and in each field is then the median value of the jackknife-resampled measurements, 
\begin{equation}
\omega_{\rm measure}(\theta) = { \underset{i}{\mathrm{med}} \: \:\omega(\theta_i)}, 
\end{equation}
with uncertainty given by Eq.~\ref{eq:jackknifeError}.

\subsubsection{The Integral Constraint}\label{sec:IC}

Estimation of $\omega(\theta)$ requires an estimate of the background galaxy density, which has to be obtained from the data sample itself.  Because the area of the survey, $\Omega$, is limited, this results in a bias in the measured correlation function, and this bias needs to be corrected.

The bias arises because the number of pairs within  the angular range $[\theta-\Delta\theta/2,\theta+\Delta\theta/2)$ is given by
\begin{equation}
N=n\left(\frac{\delta\Omega_{1}}{\Omega}\frac{\delta\Omega_{2}}{\Omega}\right)[1+\omega(\theta)].\label{eq:integr_support}
\end{equation}
Doubly integrating the quantity given by Eq.~\ref{eq:integr_support} over the solid angles $\Omega_{1}$ and $\Omega_{2}$ for the total survey area gives the total number of unique data-data pairs. However, this method gives an overestimation of the mean density due to positive correlation between galaxies at small separations \citep{1994A&A...282..353I}, which is balanced by negative correlation at larger separations. The magnitude of that effect depends on both the field size and the clustering strength.  We correct for this bias using the standard ``integral constraint'' approach, as follows. 

Let $\omega_{\rm measure}$ be the measured correlation function, which is related to the actual correlation function $\omega_{true}$ (e.g, \citealt{Sato2014}) by
\[
1+\omega_{\rm measure}(\theta)=f(1+\omega_{\rm true}(\theta)),
\]
 where $f$ is a scaling factor defined later. Using Equation \ref{eq:integr_support} and the constraint
\[
N=\iint n\left(\frac{\delta\Omega_{1}}{\Omega}\frac{\delta\Omega_{2}}{\Omega}\right)f[1+\omega(\theta)],
\]
gives $f=1/(1+IC)$, where $IC$ is the so-called ``integral constraint'', which corrects for the bias mentioned above.  The negative offset is given by integrating the assumed true $\omega(\theta)$
over the field $\Omega$ \citep{PeeblesBook},
\[
IC=\frac{1}{\Omega^{2}}\iint\omega(\theta)d\Omega_{1}d\Omega_{2},
\]
where $\Omega$ corresponds to the solid angle of the survey. In practice, the above integral is well approximated using
\[
IC=\frac{\sum RR(\theta)A_{\omega}\theta^{-\beta}}{\sum RR(\theta)},
\]  \citep{1999MNRAS.307..703R,1994A&A...282..353I}, which includes the random-random pair counts, $RR$.  The result is added to the measured value $\omega_{\rm measure}(\theta)$ to obtain the true value $\omega_{true}(\theta)$, namely
\begin{equation}
\omega_{\rm true}(\theta)\approx\omega_{\rm measure}(\theta)+IC.
\end{equation}

It is well known that the two-point angular correlation function is well approximated by a power law \citep{PeeblesBook} of the form
\begin{equation}
\label{eq:powerlaw}
\omega(\theta)=A\theta^{1-\gamma}.
\end{equation}
Assuming the above power law form in Eq.~\ref{eq:powerlaw}, the data are fit using a non-linear least-squares fit to estimate the parameters $A_{\omega}$ and $\gamma$ to quantify the strength of clustering. For $\omega_{\rm true}(\theta)=A_{\omega}\theta^{1-\gamma}$, the estimated correlation function is given by $\omega_{\rm measure}(\theta)=A_{\omega}(\theta^{1-\gamma}-C)$, where $C=\frac{IC}{A_{\omega}}$. The value of IC is found to range from 0.06 to 0.08 for the Deep fields and 0.04 to 0.06 for the Wide fields, and we corrected our clustering measurements using these IC values.

\subsection{Results}\label{sec:2Dresults}

\begin{figure}
\includegraphics[width=8.5cm,keepaspectratio]{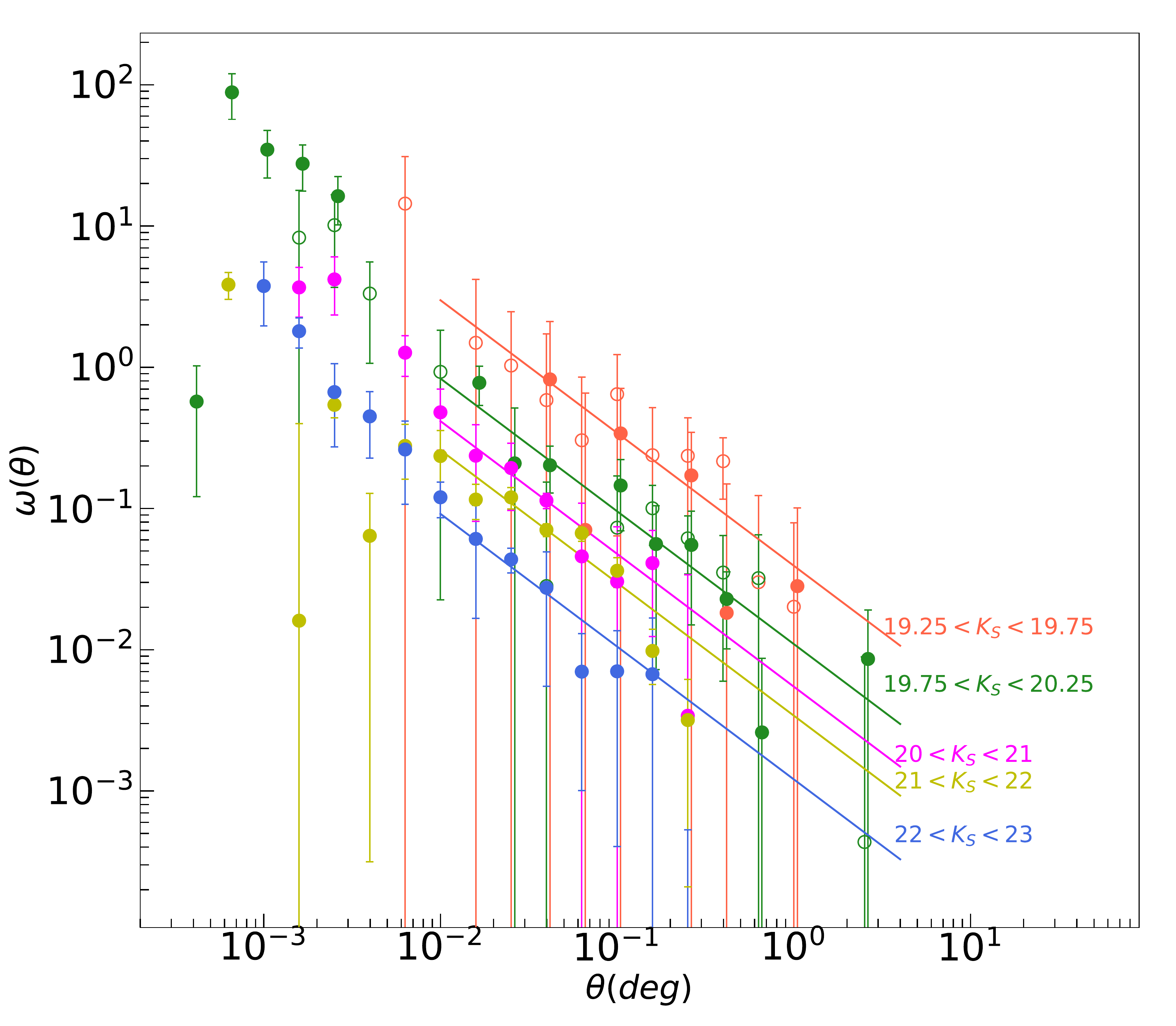}
\protect\caption[Angular correlation function of \PEgzK\ galaxies as a function of $K_{s}$-band magnitude.]{\label{fig:-The-angular-correlation} The angular correlation function of \PEgzK\ galaxies as a function of $K_{s}$-band magnitude. The open and filled circles for the Wide fields represent W1 and W4 field measurements, respectively.  The magnitude intervals are $19.25<K_{s}<19.75$, and $19.75<K_{s}<20.25$ for the Wide fields and $20<K_{s}<21$, $21<K_{s}<22$, $22<K_{s}<23$ for the Deep fields.  Solid lines show fits to the data done at large angular scales ($\theta>10^{-2}$deg, where the one-halo term is negligible) and with $\gamma=1.92$ fixed.
}\end{figure}

Figure \ref{fig:-The-angular-correlation} summarizes our clustering
measurements, corrected by the IC procedure described above, as a function of \Ks\ magnitude. We use logarithmic angular binning of $\Delta\log\theta=0.2$, where $\theta$ is in degrees, to provide adequate sampling at small scales and to avoid excessively fine sampling and poor signal-to-noise ratios at large scales. The upper limit for $\theta$ has to be smaller than the field size and the lower limit  is set by the lack of galaxy pairs at small separations. Therefore, in the Deep fields $\omega(\theta)$ is computed over $-3.5 < \log(\theta) < -0.5$;  in the Wide fields the range is $-3.5 < \log(\theta) < 0.5$.

\begin{table*}
\label{tab:summary}
\begin{centering}
\begin{tabular}{|l|c|c|c|c|c|c|c}
\hline
Fields & $K_{s}${[}AB mag{]} & $\log(M_{\star}$/$M_{\odot}$) & $N_{D}$ & $A_{\omega}/10^{-3}$(deg)$^{1-\gamma}$ & arcmin ($10^{-2}$)  & $r{}_{0,AO} $($h^{-1}$Mpc) & $r_{0,B}$($h^{-1}$Mpc)\tabularnewline
\hline
Wide & 19.25-19.75 & 11.49 & 132+71   & 39.32$\pm$ 6.78 & 170.02 $\pm$ 29.32&  29.77 $\pm$2.75 & 20.49$\pm$1.89  \tabularnewline
         & 19.75-20.25 & 11.32 & 675+434 & 10.97$\pm$ 2.09 & 47.44 $\pm$ 9.03 &  15.31$\pm$2.12 & 10.54$\pm$1.46  \tabularnewline
Deep & 20-21 & 11.15 & 841                  & 4.89$\pm$ 0.49 & 21.15$\pm$ 2.12 & 10.05$\pm$0.47 & 7.34$\pm$0.34 \tabularnewline
         & 21-22 & 10.80 & 2282                & 2.84 $\pm$ 0.42 & 12.28 $\pm$ 1.83 &  7.57$\pm$0.49 & 5.73$\pm$0.37  \tabularnewline
         & 22-23 & 10.45 & 1881                & 1.79 $\pm$ 0.40 & 7.74 $\pm$ 1.73 & 5.95$\pm$1.03 & 3.34$\pm$0.58  \tabularnewline
\hline
\end{tabular}
\par\end{centering}
\caption{The clustering amplitudes $A_{\omega}$ and $r_0$ for our $gzK_s$-selected passive galaxies as a function of $K_{s}$-magnitude bin or, equivalently, stellar mass (with stellar computed using Eq.~\ref{eq:KtoMass}).   For the Wide fields the reported number of objects, $N_D$, is given for the two fields separately (W1+W4), while for the Deep fields the  number is the sum over the four fields. The estimated clustering lengths are measured over the angular separation range $0.01{}^{\circ}<\theta<0.32^{\circ}$ for Deep fields and $0.013{}^{\circ}<\theta<0.631^{\circ}$ for the Wide fields, in both cases with power-law slope fixed at $\gamma=1.92$.  The two $r_0$ columns represent the 3-D correlation lengths for the two different redshift distributions we considered:  $r_{0,AO}$ for the $N(z)$ given in \citet{LARGE1} and $r_{0,B}$ for that in \citet{0004-637X-681-2-1099}. Our preferred $r_0$ values are those computed with the redshift distribution of \citet{LARGE1}. }
\end{table*}

The angular correlation measurements for the two Wide fields,
W1 and W4 were kept separate and thus treated as independent measurements. For the Deep fields, the measurements from the four different independent fields are combined using a weighted mean. Assuming the points come from the same parent populations with the same mean, but different standard deviations, the weighted average of the angular correlation function is given by
\[
\bar{\omega}=\frac{\sum_{i}(\omega_{i}/\sigma_{i}^{2})}{\sum_{i}(1/\sigma_{i}^{2})},
\]
where each data point $\omega_{i}$ is inverse variance weighted. With $w=1/\sigma_{\omega}^{2}$ as the weight, the uncertainty of the mean $\sigma$ is given by
\[
\sigma^{2}=\frac{\sum_{i}(1/\sigma_{i}^{2})}{(\sum_{i}(1/\sigma_{i}^{2}))^{2}}=\frac{1}{\sum_{i}(1/\sigma_{i}^{2})},
\]
and the variance of the weighted mean is
\[
\sigma^{2}=\frac{\sum_{i}w_{i}(\omega_{i}-\omega)^{2}}{(\sum_{i}w_{i})}\times\frac{1}{N-1},
\]
where N=4 is the number of fields.

The $K_{s}$-band magnitude gives an approximate measure of stellar masses of UMPEGs \citep[][see also Eq.~\ref{eq:KtoMass}]{LARGE1} as the rest-frame $\sim$8500\AA\ light from passive galaxies at $z\sim1.6$ is dominated by the long-lived low-mass stars that contain most of the stellar mass in a stellar population.
We will discuss the stellar mass dependence of clustering in more detail in Sec.~\ref{sec:SHMR} but in this Section we focus on the more direct \Ks-magnitude dependence of clustering, while noting that the two quantities, mass and magnitude, are related. Here, we divided our sample into subsamples based on $K_{s}$-band luminosity.  In the Wide fields, we divided the \PEgzK\ sample into two sub-samples of bin size 0.5 mag: $19.25<K_{s}<19.75$, and $19.75<K_{s}<20.25$.  In the Deep fields the population is divided into three subsamples with bin-size of 1.0 mag: $20<K_{s}<21$, $21<K_{s}<22$, $22<K_{s}<23$. 

We fitted the measurements with power laws of the form given by Eq.~\ref{eq:powerlaw}, and the results are shown with solid lines in  Fig.~\ref{fig:-The-angular-correlation}, while the corresponding best-fit parameter values are listed in Table~1. For the Wide fields, for which the W1 and W4 measurements were kept separate, the $\omega(\theta)$ values from the two fields were treated as independent measurements and fitted simultaneously. For the Deep fields the fits were done to the combined values from the four fields.  The fits were performed over angular scales of $0.01^{\circ}$ to $0.32^{\circ}$ for the Deep fields and $0.013^{\circ}$ to $0.631^{\circ}$ for the Wide fields. The power law index for the fainter  ($22<K_{s}<23$) passive galaxies in the Deep fields is found to be $\gamma=1.92\pm0.12$. The other, brighter sub-samples were fitted allowing the power-law amplitude to vary while keeping $\gamma$ fixed at that 1.92.

We clearly found a positive correlation function signal for the passive galaxies in both Deep and Wide fields and in all magnitude ranges studied, with an angular dependence consistent with slope $\gamma=1.92$.  This is in agreement with the results of \citet{Sato2014} who also found $\gamma$ to be 1.92 (with the same Deep dataset that we use) for \gzK-selected passive galaxies, although we note that in the present paper we probe the previously unstudied ultra-massive regime.  We note that, in contrast, \citet{McCracken2010} found the best fitting slope $\gamma$ for their passive $BzK$-selected galaxies to be $\gamma\sim2.3$.

As can be seen in Fig.~\ref{fig:-The-angular-correlation}, the correlation function deviates from the power-law at small angular scales, and this deviation was also found for \PEgzK\ galaxies in CFHTLS Wide fields by \citet{Sato2014}. This deviation is due to the so-called one-halo term which is the contribution from galaxy pairs residing within the same dark halo and is related to dark matter halo substructure \citep{0004-637X-575-2-587}. The power law on large scales is due to the two-halo term, which represents the clustering of galaxies that reside in distinct halos and which dominates on scales larger than the virial radius of a typical halo. As was mentioned above, our fits to the  clustering measurements are done at large $\theta$ in order to measure only the clustering of galaxies residing in the distinct halos.  Because our UMPEGs are very massive and don't have similar- or larger-mass companions, our UMPEG auto-correlation measurement can thus be expected to measure the clustering of the dark matter halos in which UMPEGs are the central galaxies.  We note that there are no  $19.25<K_{s}<19.75$ UMPEG pairs in our catalog at separations smaller than 57.06 arcsec, which is consistent with this expectation. 

It is clear in Figure~\ref{fig:-The-angular-correlation} that the clustering strength increases monotonically with \PEgzK\ galaxy brightness in the \Ks-band.  This is also shown in Fig.~\ref{fig:clustering_vs_mag}, which shows the values of the 2D clustering amplitude, $A(\omega)$, as a function of \Ks-band magnitude.  This trend is consistent with, but clearer than, previous studies  \citep{McCracken2010, Sato2014, doi:10.1093/mnras/stv1927}).  Moreover, our sample extends over a much wider magnitude range than these previous studies, and probes extremely bright, previously unstudied ultra-massive objects.


\begin{figure}\label{fig:clustering_vs_mag}
\begin{centering}
\includegraphics[width=8.5cm,keepaspectratio]{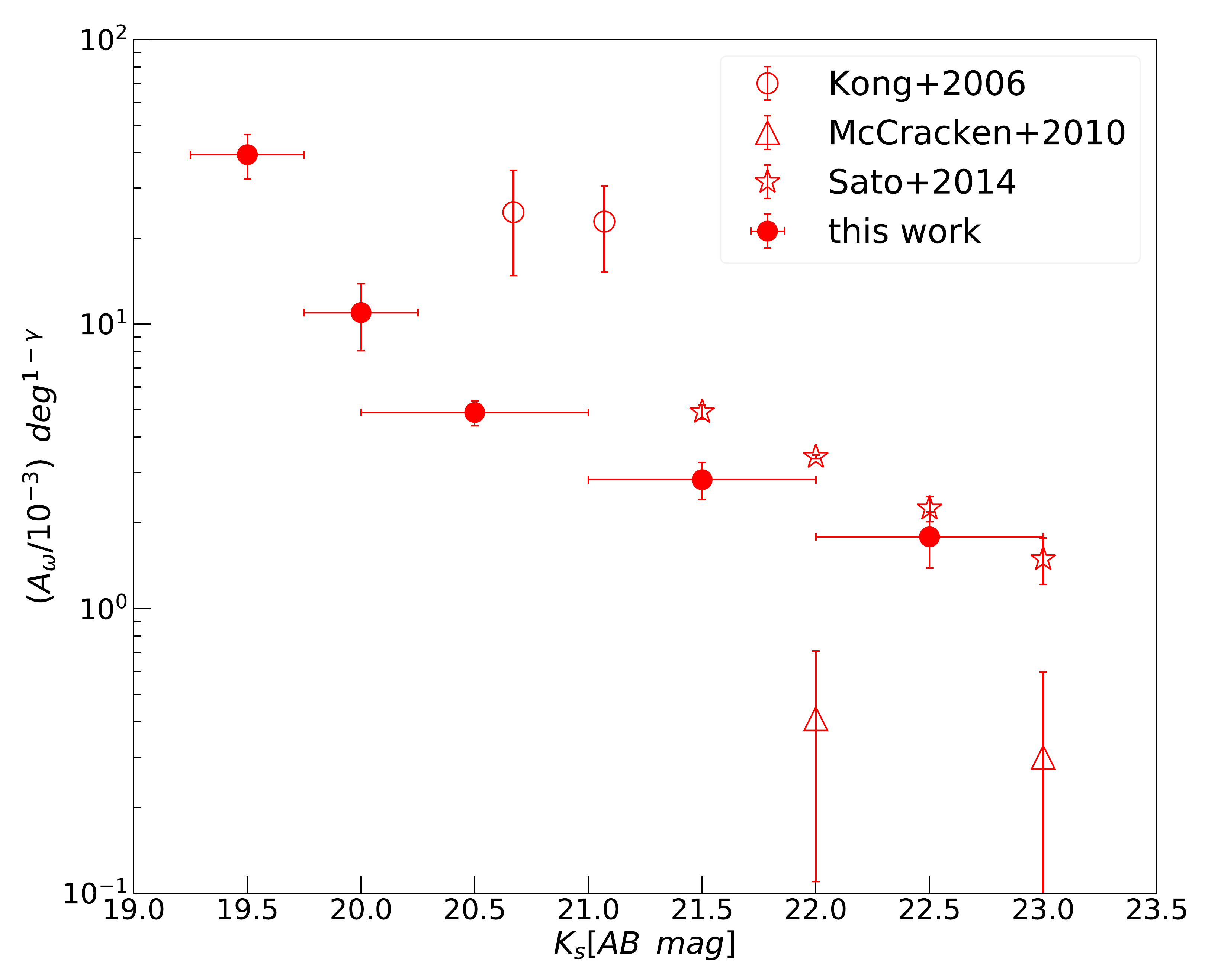}
\par\end{centering}
\protect\caption[Clustering amplitude for \gzK-selected passively evolving galaxies as a function of $K_{s}$ magnitude.]{\label{fig:clustering_vs_mag} Clustering amplitude for \gzK-selected (or $BzK$-selected) passively evolving galaxies as a function of $K_{s}$ magnitude. The horizontal bars indicate not uncertainties but the $K_{s}$ magnitude intervals that define our  subsamples, while vertical errorbars reflect uncertainties from jackknife resampling of the dataset.  The measurements were done with $\gamma$ fixed at 1.92.}
\end{figure}


\section{The Spatial Correlation Function}\label{sec:Spatial-Correlation-Function}

The two-dimensional galaxy correlation function, $\omega(\theta)$,  is the projection of the three-dimensional clustering, $\xi(r)$, which is the underlying physical relation. The spatial correlation function $\xi(r)$ is defined analogously to the definition of $\omega(\theta)$ provided by Eq.~\ref{eq:angular correlation}.  Considering two infinitesimally thin shells centered on two objects, located at $r_{1}$ and $r_{2}$,  $\xi(r)$ is defined by the joint probability $dP(r)$ of finding two objects within volume elements $dV_{1}$ and $dV_{2}$, at a separation $r=r_{1}-r_{2}$ such that
\[
dP(r)=n[1+\xi(r)]dV_{1}dV_{2},
\]
where n is now the space density of objects. The spatial correlation function can be described as a power law of the form
\[
\xi(r)=\left(\frac{r}{r_{o}}\right)^{-\gamma},
\]
where $r$ is the co-moving distance between the two points, $r_{o}$
is the characteristic correlation length, and $\gamma$ is the slope derived from the angular correlation measurements.

We do not measure $\xi(r)$ directly, but can infer it from the angular correlation function, $\omega(\theta)$, and the redshift distribution of our galaxy sample, $N(z)$, by means of the inverse Limber transformation \citep{Limber1953}. 

\subsection{Limber Inversion}\label{sec: limber inversion}

The de-projection of the angular correlation function is done using the Limber inversion as follows. The amplitudes of the power law representations of the angular and spatial correlation functions are related by the  equation\begin{equation}
A_{\omega}=\frac{H_{\gamma}r_{0}^{\gamma}\int_{0}^{\infty}F(z)r_{c}^{1-\gamma}(z)N^{2}(z)E(z)dz}{(c/H_{0})[\int_{0}^{\infty}N(z)dz]^{2}}
\label{eq:limber eqn}
\end{equation}
 \citep{Limber1953, doi:10.1046/j.1365-8711.1999.02612.x}, 
where $A_{\omega}$ is the amplitude of $\omega(\theta)$, $r_{c}(z)$ is the radial co-moving distance at redshift $z$, and $H_{\gamma}$ is a factor that depends on the power-law index slope and is given by
\begin{equation}
H_{\gamma}=\Gamma(1/2)\frac{[\Gamma(\gamma-1)/2]}{\Gamma(\gamma/2)}.
\end{equation}
Here $E(z)$ is a cosmology-dependent expression given by
\begin{equation}
E(z)\equiv\sqrt{\Omega_{m}(1+z)^{3}+\Omega_{k}(1+z)^{2}+\Omega_{\Lambda}},\label{eq:cosmo_expression}
\end{equation}
where $\Omega_{m}$ is the matter density parameter,  $\Omega_{\Lambda}$is the cosmological constant, and the curvature of space is characterized by $\Omega_{k}=1-\Omega_{m}-\Omega_{\Lambda}$. In Eq.~\ref{eq:limber eqn}, $F(z)$ accounts for the redshift evolution of $\xi(r)$ and is assumed to be negligible within our samples and thus set to $F(z)=1$ (this describes the case of ``co-moving clustering'', where halo separations expand with the Universe). $N(z)$ corresponds to the redshift distribution of the studied galaxy population, which is an important quantity and is described in Section~\ref{sec:redshift distributions}. Finally, $r_c(z)$, 
 the radial co-moving distance between observer and an object at redshift $z$, is computed using the relation
\[
r_{c}(z)=D_{H}\int_{0}^{z}\frac{dz'}{E(z')}
\]
 \citep{Hogg1999}, where the function $E(z)$ is defined in Equation \ref{eq:cosmo_expression},
and $D_{H}$ is the Hubble distance given by $D_{H}\equiv c/H_{0}$.

\subsection{Redshift Distributions}\label{sec:redshift distributions}

The redshift distribution of the galaxy sample is a ciritcal ingredient of the Limber equation, Eq.~\ref{eq:limber eqn}, and can depend on the magnitude of the subsample being studied, thereby affecting the inferring spatial clustering.  The magnitude-dependent redshift distribution,  $N(m,z)$, of our passive \gzK\ galaxies was computed by \citet{LARGE1} by cross-correlating our \PEgzK\ samples in the D2 (COSMOS) field with the photometric redshift catalog of \citet{Muzzin2013},  and in the Wide fields (W1 and W4) with the catalog of \citet{refIda}. \citet{LARGE1} binned the photometric redshifts in magnitude steps of 0.5 width in $K_{s}$-band and the resulting $N(m,z)$ can be seen in the middle panel of their Fig.~6. A key feature of these redshift distributions is that they vary with magnitude: the peak redshift for fainter \PEgzK\ galaxies lies at somewhat higher redshifts ($z\sim 1.7$ for $K_s\sim22-23$) compared to that of the brighter galaxies ($z\sim 1.6$ for $K_s\sim18-19$), and the fainter galaxies also have a more pronounced high-redshift tail than their brighter counterparts. We adopt these $N(z, m)$ as our preferred redshift distributions for computing the spatial clustering of corresponding \PEgzK\ samples using Eq.~ \ref{eq:limber eqn}.

While we prefer the $N(z, m)$ derived by \citet{LARGE1} as described above, we also computed values of $r_{0}$ assuming the redshift distribution given by \cite{0004-637X-681-2-1099}. This assumes that the \PEgzK\ redshift distribution is a simple magnitude-independent Gaussian centred at $z=1.58$ and with width $\sigma_{z}=0.17$. Using this alternative redshift distribution gives $r_0$ values that are $\sim$2/3 times those we obtained with our preferred, magnitude-dependent redshift distribution from \cite{LARGE1}. This difference can be viewed as an indication of the systematic uncertainty in our  (and other authors') $r_0$ measurements, but ultimately we prefer the $r_0$ values we obtained with the more accurate redshift distribution given in \cite{LARGE1}.

\subsection{Estimating the Correlation Length}

\begin{figure}
\begin{centering}
\includegraphics[width=8.8cm,keepaspectratio]{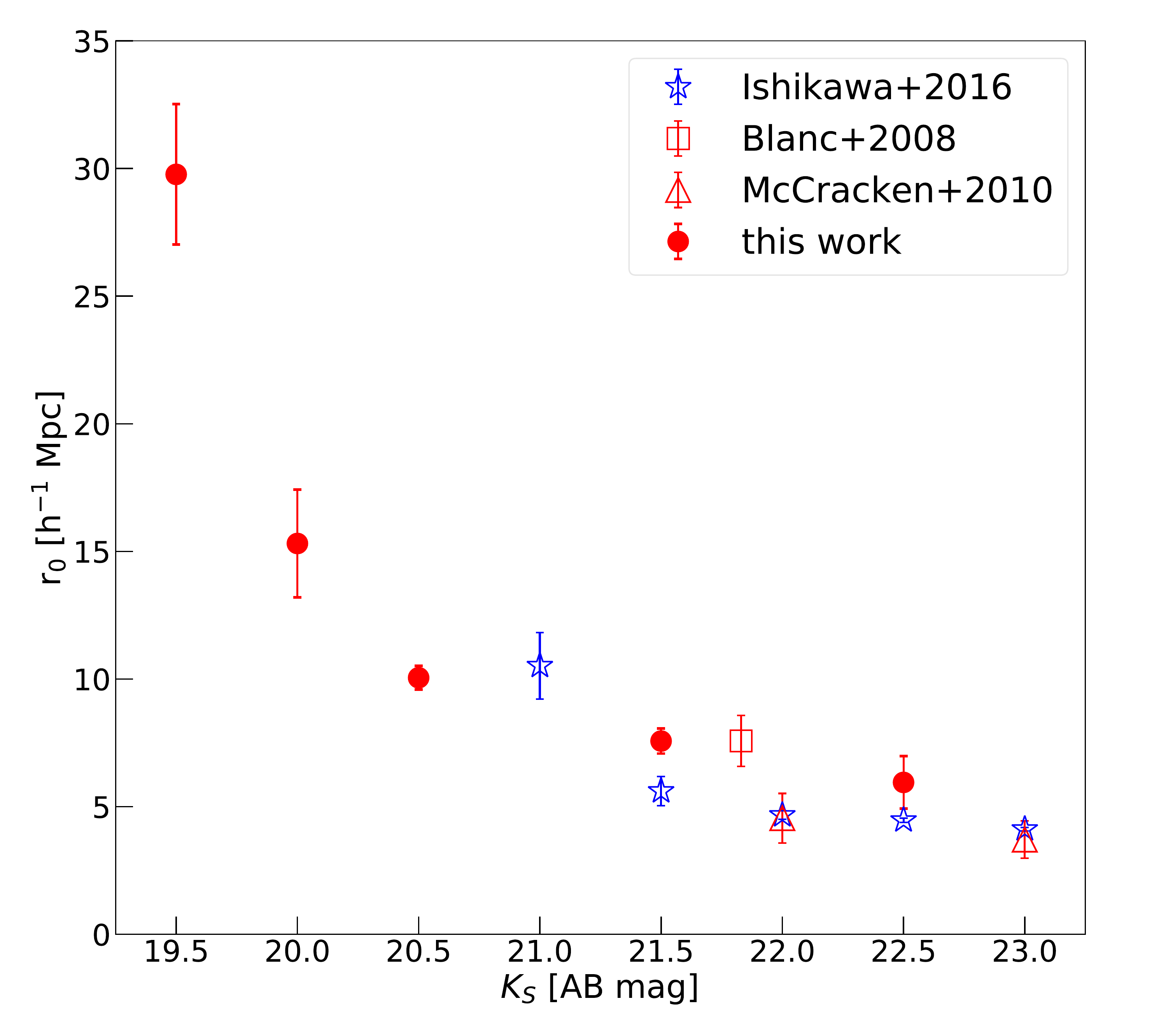}
\par\end{centering}
\protect\caption{\label{fig:ro_comparison}Comparison of the correlation lengths of our \PEgzK\ galaxies (red filled circles) with previous studies of $BzK$-selected galaxies. Star-forming $BzK$ galaxies are shown as blue points while red symbols show passive galaxies. All correlation lengths are in units of $h^{-1}$Mpc, where $h=0.7$. Compared to previous studies, our results extend to much brighter but extremely rare passive galaxies at \K$<$21.
}
\end{figure}

Table 1 and Figure \ref{fig:ro_comparison}
summarise the values calculated for the correlation length $r_{0}$ measured within our \K-magnitude selected samples using the Limber inversion. The uncertainties in $r_{0}$ given in Table~1 combine two factors: (1) the uncertainty in the measurement of $A_{\omega}$, and (2) the uncertainty in the redshift distribution, $N(z)$. The two different sets of $r_{0}$ values in Table~1 are derived from the two different redshift distributions: these of \citet{LARGE1} and \citet{0004-637X-681-2-1099}, as described in Sec.~\ref{sec:redshift distributions}. 

The uncertainties in Table~1 include only uncertainties  due to statistical errors. For uncertainties in the redshift distribution, systematic errors are expected to be larger than random errors. To see the effect of redshift distribution -- which is a systematic uncertainty -- on the estimation of $r_{0}$, we used two different redshift distributions to calculate the spatial correlation lengths for each $K_{s}$-magnitude selected sample. Here, the correlation length is affected by the median redshift and the width of the redshift distribution \citep{McCracken2010} -- a larger width in the redshift distribution implies that projection effects are stronger and would result in a larger value of $r_{0}$ for a given time or underlying clustering.


It is clear in Table~1 that the two different redshift distributions give different $r_{0}$ results. Nevertheless, in both cases, the correlation lengths increase with the increase in \K\ brightness. UMPEGs have larger correlation lengths compared to the fainter \PEgzK\ galaxies, indicating that they cluster more strongly. We note again that we consider the $r_{0, AO}$ values to be more reliable of the two, since they are derived with a more realistic set of redshift distributions. For this reason, we will use these correlation length values in all that follows. 

Figure~\ref{fig:ro_comparison} shows a comparison of our $r_{0}$ measurements with those of previous studies. For the fainter passive galaxies, the clustering strength agrees with the results for $BzK$-selected passive galaxies at similar \Ks\ magnitudes reported by \citet{0004-637X-681-2-1099} and \citet{McCracken2010}. Our results also show that the intermediate-brightness \PEgzK\ galaxies have clustering comparable to that of the star-forming galaxies of similar \Ks\ magnitude \citep{doi:10.1093/mnras/stv1927}.   However, in contrast to  previous studies, the clustering measurements in our work also extend to much brighter passive galaxies. The fact that our very luminous passive galaxies cluster more strongly than the fainter $BzK$ galaxies (both passive and star-forming) suggests that they reside in much more massive dark matter halos.

\section{Masses of dark matter halos}\label{sec:haloMasses}

In the $\Lambda$CDM model, the clustering of DM halos is well understood \citep[e.g., ][]{doi:10.1046/j.1365-8711.2002.05723.x}: the halos cluster in such a way that the most massive DM halos have larger clustering strength as measured by the correlation function. This trend arises because the DM halos (and the galaxies inside them) form from small perturbations in the early Universe which grow with time. Here, high mass halos are formed in regions with strong, positive perturbations
on even larger scales \citep{1984ApJ...284L...9K}. The large-scale collapse accelerates the collapse of the smaller halos, causing an excess of these halos in the general neighbourhood and hence, strong clustering of massive halos. In contrast, large scale perturbations are not needed to form the low mass halos and hence low mass halos have weaker clustering.

Because clustering amplitude is a monotonically increasing function of halo mass, we can use the observed clustering of our \PEgzK\ galaxies, compared with that of DM halos in a $\Lambda$CDM N-body simulation, to identify the dark matter halo masses of our galaxies.  With this goal in mind, in this Section we match the clustering strengths we measured in Sec.~\ref{sec:Spatial-Correlation-Function} to the clustering strengths of dark matter halos in the Millennium {\it XXL} simulation \citep[\MXXL;][]{doi:10.1111/j.1365-2966.2012.21830.x}.

\subsection{Brief Review of the Millennium XXL Simulation\label{sec:Halos}}

The \MXXL\ simulation  \citep{doi:10.1111/j.1365-2966.2012.21830.x} is a very large, high-resolution cosmological dark matter N-body simulation that  greatly extends the previous Millennium and Millennium-II simulations \citep{2005Natur.435..629S,doi:10.1111/j.1365-2966.2009.15191.x}. The simulation follows the non-linear evolution of $6720^{3}$=303,464,448,000 dark-matter particles with masses $6.2\times10^{9}h^{-1}M_{\odot}$ each, distributed within a cubic box of comoving length 3 $h^{-1}$ Gpc, which is equivalent to the volume of the whole observable Universe to redshift $z$=0.72.  This mass resolution is sufficient to identify host dark matter halos of galaxies with stellar masses greater than $1.5\times10^{10}$$M_{\odot}$ \citep{doi:10.1111/j.1365-2966.2005.09879.x}, while the very large volume is large enough to contain very rare, very massive halos (the most massive halo at $z=0$ has $M_{FoF}=10^{15.95}M_{\odot}$). Because of their very large $r_0$ (Sec.~\ref{sec:Spatial-Correlation-Function}), our UMPEGs can be expected to reside in very massive halos, and for this reason the large-volume \MXXL\ simulation is well-suited for our purposes.

\MXXL\ adopts the $\Lambda$CDM cosmology with WMAP-1 cosmological parameters with the total matter density $\Omega_{m, 0}=0.25$ and cosmological constant $\Omega_{\Lambda, 0}=0.75$; the RMS linear density fluctuation in 10.96 Mpc spheres, extrapolated to the present epoch, is $\sigma_{8}=0.9$; and $H_{0}=0.73$ km s$^{-1}$ Mpc$^{-1}$. While these cosmological parameters are somewhat different than those we used in estimating our $r_0$ values from the observational data, they are sufficiently similar that we do not need to make any adjustments in our analysis. 

The simulation follows the gravitational growth traced by its
DM particles and stores it as DM particle positions at 64 discrete time snapshots. The initial conditions are set at a starting redshift of $z=127$ and the simulation evolves to $z=0$ with 63 outputs corresponding to various redshifts.  At each snapshot, groups of more than 20 particles are identified as dark matter halos using a Friends-of-Friends (FoF) algorithm \citep{1985ApJ...292..371D}. Following this step, the SUBFIND algorithm \citep{doi:10.1046/j.1365-8711.2001.04912.x} finds gravitationally-bound subhalos within each FoF halo. The mass of the halo is defined as the conventional virial mass of a halo, which is $M_{200}=M(r<r_{200})$, the mass contained within a sphere of a radius that encloses a mean density that is 200 times the critical density of the Universe.

\subsection{Clustering of Dark Matter Halos at $z\sim1.6$}

We used the halo catalog from snapshot=36  of the {\it MXXL} simulation, which corresponds to  $z\sim1.6$,  the peak redshift of the UMPEG redshift distribution. The spatial correlation function of the DM halos is a function of halo mass \citep*{doi:10.1093/mnras/282.2.347}, and for this reason we study halo clustering of halos within specific mass ranges.  Specifically, we divided the \MXXL\ halos into eleven logarithmic halo mass bins of width $\Delta\!\log(M_{200})$=0.2 that span $12.1<\log(M_{200})<14.3$; here $M_{200}$ is in units of $h^{-1}M_{\odot}$. For every halo mass bin, we then  measure the two-point correlation function of the dark matter halos, given by
\[
\xi(r)=\frac{DD(r)-2DR(r)+RR(r)}{RR(r)}.
\]
This equation is similar to  Eq.~\ref{eq:LS}, but here $DD(r)$ is the unique number of halo pairs in the simulation with separations between $r-\delta r<r<r+\delta r$, $DR(r)$ refers to the number of pairs within the same separations between the halo catalog and a random distribution of positions, and $RR(r)$ refers to number of random-random pairs within the same range.

\begin{figure}
\begin{centering}
\includegraphics[width=8.5cm,keepaspectratio]{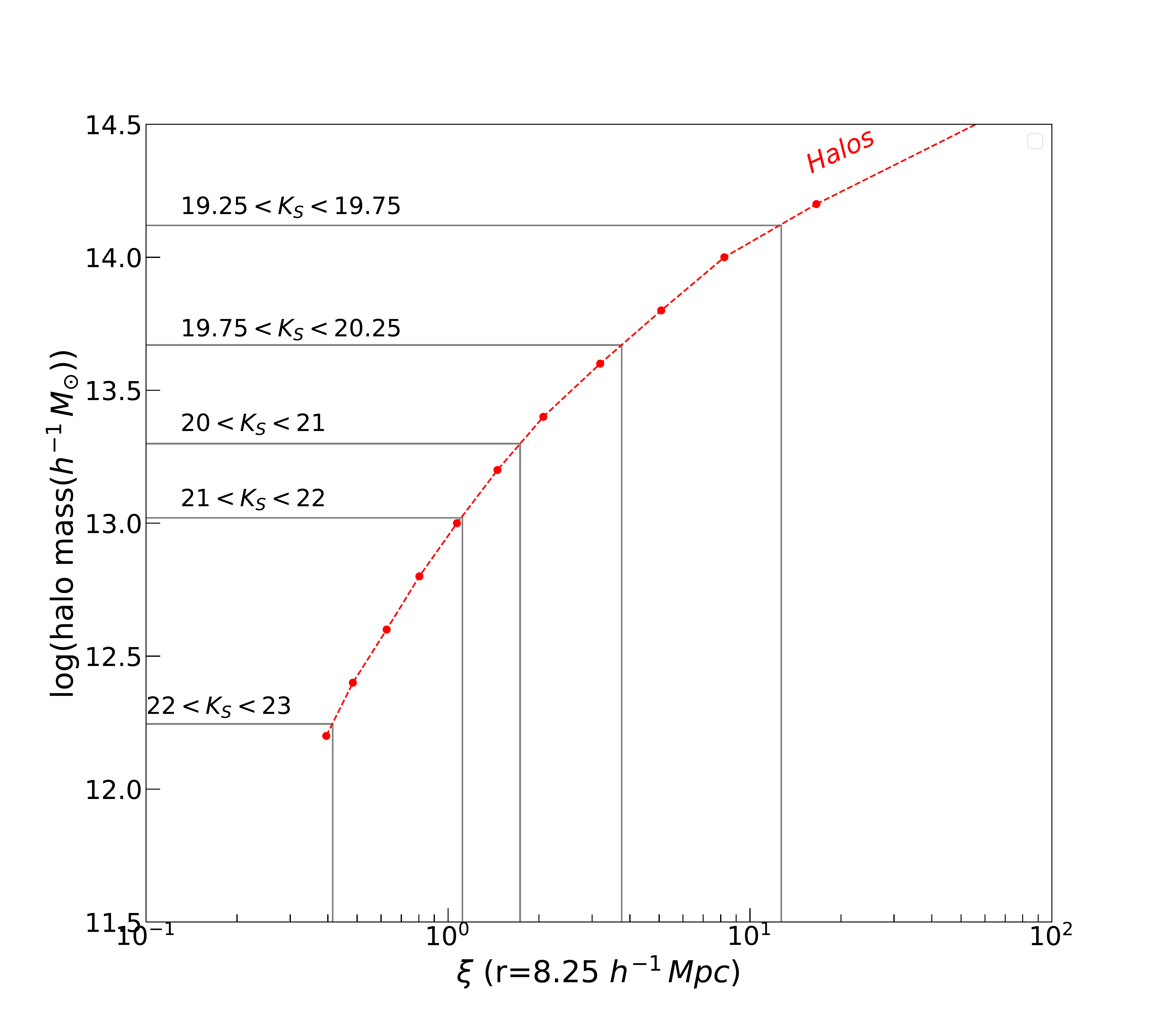}
\par\end{centering}
\caption{\label{fig:Halo-mass-xi}Halo mass as a function of the correlation function of dark matter halos in the $z = 1.6$ snapshot of the {\it MXXL} simulation; the same fixed slope $\gamma$ was used for these measurements as in the measurements of the \PEgzK\ galaxies.  Red points show the  correlation function at a fixed spatial value $r=8.25\mbox{ }h^{-1}$Mpc for different halo masses. The dashed red line is an interpolation between the {\it MXXL} data points.  The vertical solid gray lines correspond to our measurements of the spatial correlation function of the \PEgzK\ galaxies binned according to their $K_{s}$-band brightness, as indicated in the Figure. The horizontal gray lines show the corresponding halo masses that result from applying the {\it MXXL} relation to our clustering observations.}
\end{figure}

Next, we compared the halo mass values of the eleven bins with their measured $\xi(r)$ values at $r$=8.25 $h^{-1}$Mpc.  These are shown using red points in Fig.~\ref{fig:Halo-mass-xi}; while the dashed red line in  Fig.~\ref{fig:Halo-mass-xi} shows a piecewise interpolation between the points.  The procedure is done at $r$=8.25 $h^{-1}$Mpc to ensure a good number of halo-halo pairs and to avoid the contribution of subhalos within larger halos (i.e., the one-halo term).  As expected, halo mass and clustering strength correlate monotonically.

\begin{figure*}
\noindent \begin{centering}
\includegraphics[width=17.5cm,keepaspectratio]{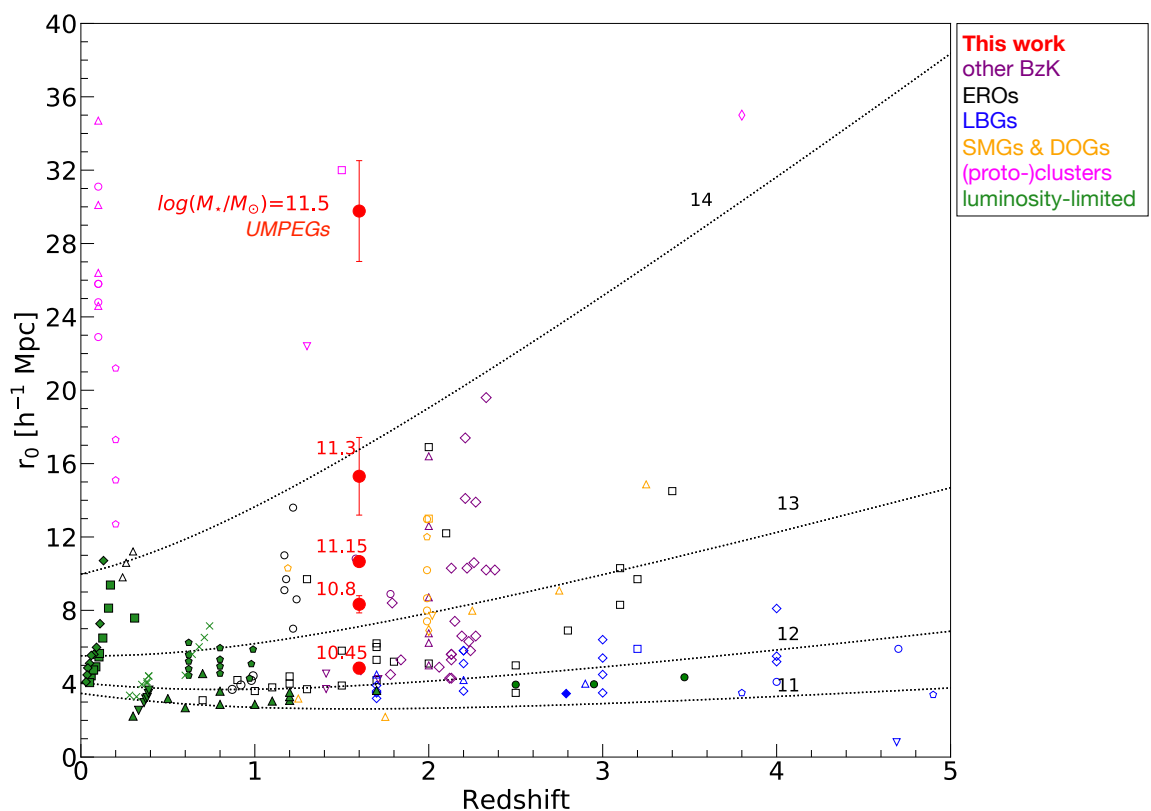}
\par\end{centering}
\protect\caption[]{The comoving correlation length $r_{0}$ of the passive galaxies from our sample (red points) in comparison with other populations of objects at a range of redshifts.  Different colours indicate different types of objects or selection techniques, with open symbols showing measurements based on photometric data, while filled symbols are for measurements from spectroscopic data (except for our points, which are photometric).  Except for our points, errorbars are omitted for clarity. Red numbers indicate stellar masses, in logarithmic units, for our subsamples.   Black curves, labelled in logarithmic units of solar mass, show the dependence of $r_{0}$ for halos of fixed minimum mass as a function of redshift from the modified Press-Schechter analysis of \citet*{doi:10.1046/j.1365-8711.2002.05723.x}. 

\hspace{1cm}The $r_0$ values plotted are from the compilation of \citet{durk} but with some additions and are as follows: 
Purple: other BzK galaxies ($\!$
open circles \textendash{} \citealt{0004-637X-681-2-1099};
open triangles \textendash{} \citealt{2010MNRAS.407.1212H};
open reversed triangle \textendash{} \citealt{McCracken2010};
open diamonds \textendash{} \citealt{refLin}
$\!$).
Blue: LBG galaxies ($\!$
open squares \textendash{} \citealt{refIdfou};
open circles \textendash{} \citealt{refouchi};
open triangles \textendash{} \citealt{Adelberger2005};
open reversed triangles \textendash{} \citealt{Kashikawa2006};
open diamonds \textendash{} \citealt{0004-637X-737-2-92};
filled diamonds \textendash{} \citealt{doi:10.1093/mnras/sts639};
open pentagon \textendash{} \citealt{0004-637X-793-1-17}
$\!$).
Green: galaxy samples from surveys limited in luminosity ($\!$
filled squares \textendash{} \citealt{doi:10.1046/j.1365-8711.2002.05348.x};
open circles \textendash{} \citealt{ref_coil}; filled triangles \textendash{}
\citealt{refIdle}; filled reversed triangles \textendash{} \citealt{refIdpol},
filled diamonds \textendash{} \citealt{Zehavi2011}; filled pentagons
\textendash{} \citealt{refIdmar}; crosses \textendash{} \citealt{0004-637X-784-2-128}
$\!$).
Red: EROs or massive red galaxies (
open squares \textendash{} \citealt{0004-637X-588-1-50};
filled squares \textendash{} \citealt{Zehavi2011};
open circles \textendash{}
\citealt{ref_brown}
$\!$).
Orange: SMGs or DOGs ($\!$
open diamonds --- \citealt{Blain2004};
open squares --- \citealt{Weiss2009};
open inverted triangles --- \citealt{Hickox2012};
open triangles --- \citealt{Wilkinson2017};
open circles --- \citealt{Brodwin2008};
open pentagons --- \citealt{Toba2017}
$\!$).
Magenta: clusters or protoclusters($\!$
open diamond --- \citealt{Toshikawa2018};
open square --- \citealt{Rettura2014};
open inverted triangle --- \citealt{Papovich2008});
open triangles --- \citealt{Abadi1998};
open circles --- \citealt{Collins2000};
open pentagons --- \citealt{Bahcall2003}).
}
\label{fig:r0-comparison}
\end{figure*}

\subsection{Halo Masses of \PEgzK\ Galaxies}\label{sec:haloMasses}

Also plotted in
Fig.~\ref{fig:Halo-mass-xi},
using vertical gray lines are the $\xi(r=8.25 h^{-1}$ Mpc) values for our observed \PEgzK\ samples. Using the relation shown with red points and line, we can then relate the clustering strengths of the \PEgzK\ subsamples to the masses of halos in the simulation. It is clear that the brightest passive galaxies in the range $19.25<K{}_{s}<19.75$ reside in the most massive halos in the mass range $13.9<\log(M_{200})<14.2$, where $M_{200}$ has the units of $h^{-1}M_{\odot}$.  Fainter \PEgzK\ subsamples are associated with lower-mass halos.

We believe that our UMPEG halo mass estimates are robust.  Halo assembly bias \citep{Gao2005, Wechsler2006, Gao2007}, while potentially significant at lower halo masses,  does not appear to have a significant effect on very massive halos (see, e.g., Fig.~2 of \citealt{Gao2005}) such as those associated with the UMPEGs.  Moreover, while the presence of the one-halo term seen for the lower-mass \PEgzK\ galaxies may affect our clustering length measurements for these populations, the measurement for the UMPEGs is unlikely to be affected by this effect given the absence of the single-halo term in this population as well as the lack of UMPEG-mass companions (described in M.~Sawicki, submitted to MNRAS).  We conclude that our UMPEG halo mass estimates,  presented above, are thus robust.

\section{Discussion}\label{sec:discussion}

\subsection{Comparing UMPEGs with other Galaxy Populations}

After obtaining the correlation lengths, $r_{0}$, for our \PEgzK\ samples at $z\sim1.6,$ we put them in the context of other populations.  This comparison is shown in Fig.~\ref{fig:r0-comparison} and includes different galaxy populations as well as galaxy (proto)clusters.

The $r_{0}$ for our less massive \PEgzK\ galaxies at $z\sim1.6$ is comparable to the $r_{0}$ measured for $BzK$ galaxies and EROs at $z\sim2$ as well those for dust-obscured galaxies (DOGs) and sub-millimetre Galaxies (SMGs) at similar and higher redshifts.  However, the correlation length $r_{0}$ of the UMPEGs (\Mstars =$10^{11.5}$\Msun\ in Fig.~\ref{fig:r0-comparison}) is larger than those of other galaxy populations at similar or higher redshifts. Instead, UMPEGs have $r_0$ that is very similar to that of \zs1.5 Spitzer/IRAC-selected galaxy clusters of \citet[][$r_0=32.0 \pm7h^{-1}$Mpc, open magenta square in Fig.~\ref{fig:r0-comparison}]{Rettura2014} and consistent with the \zs1.3 clusters of \citet[][$r_0=22.4 \pm3.6h^{-1}$Mpc, open inverted magenta triangle]{Papovich2008}. It is also worth to point out here the large $r_0$ value for the $gri$-selected protocluster candidates at \zs3.8 \cite[][open magenta diamond, $r_0=35.0^{+3.0}_{-3.3} h^{-1}$Mpc]{Toshikawa2018}.  Altogether, the clustering of our UMPEGs is stronger than that of any other galaxy population, and is more consistent with the clustering of high-$z$ (proto)clusters.

Figure~\ref{fig:r0-comparison} also shows the expected clustering of halos of fixed mass as a function of redshift.  This is shown with black curves (with masses labelled in logarithmic units of solar mass) and is based on the Press-Schechter formalism \citep{PressSchechter1974} for the clustering of DM halos from \cite{doi:10.1046/j.1365-8711.2002.05723.x}.  These  models are only a first approximation as they are based on simplified assumptions of the Press \& Schechter theory, but they give an indication of the masses of the halos likely to be associated with the populations shown in Fig.~\ref{fig:r0-comparison}. Notably, these Press-Schechter masses are also consistent with the halo masses we get for our UMPEGs and other \PEgzK\ galaxies from the \MXXL\ comparisons presented in Section~\ref{sec:haloMasses}.

\subsection{Stellar Mass - Halo Mass Relation}\label{sec:SHMR}

The ratio of the stellar mass of a galaxy and the mass of its host DM halo (the so-called stellar-to-halo mass ratio, SHMR =   \Mstars$/M_{h}$) is related to the efficiency with which the galaxy can form stars and thus is of key interest in understanding galaxy formation.  With this in mind, we investigate the SHMR for our \PEgzK\ galaxies as a function of halo mass.  Here we estimate stellar masses by using the \Mstars--\Ks\ relation for \PEgzK\ galaxies calibrated on the COSMOS data of \citet{Muzzin2013}, which is given by 
\citet{LARGE1} as 
\begin{equation}\label{eq:KtoMass}
\log[M_\star/M_\odot]=-0.348K_s+18.284.
\end{equation}
We note that while SED fitting of unresolved photometry, such as that done by \citet{Muzzin2013}, can significantly underestimate stellar masses of {\it star-forming} galaxies  \citep{Sorba2015, Sorba2018, Abdurrouf2018}, such bias is not present for {\it quiescent} galaxies such as the \PEgzK\ systems we study here.  Consequently, \PEgzK\ galaxy stellar masses estimated using Eq.~\ref{eq:KtoMass} can be expected to be reasonably accurate. 

\begin{figure}
\begin{centering}
\includegraphics[width=8.5cm,keepaspectratio]{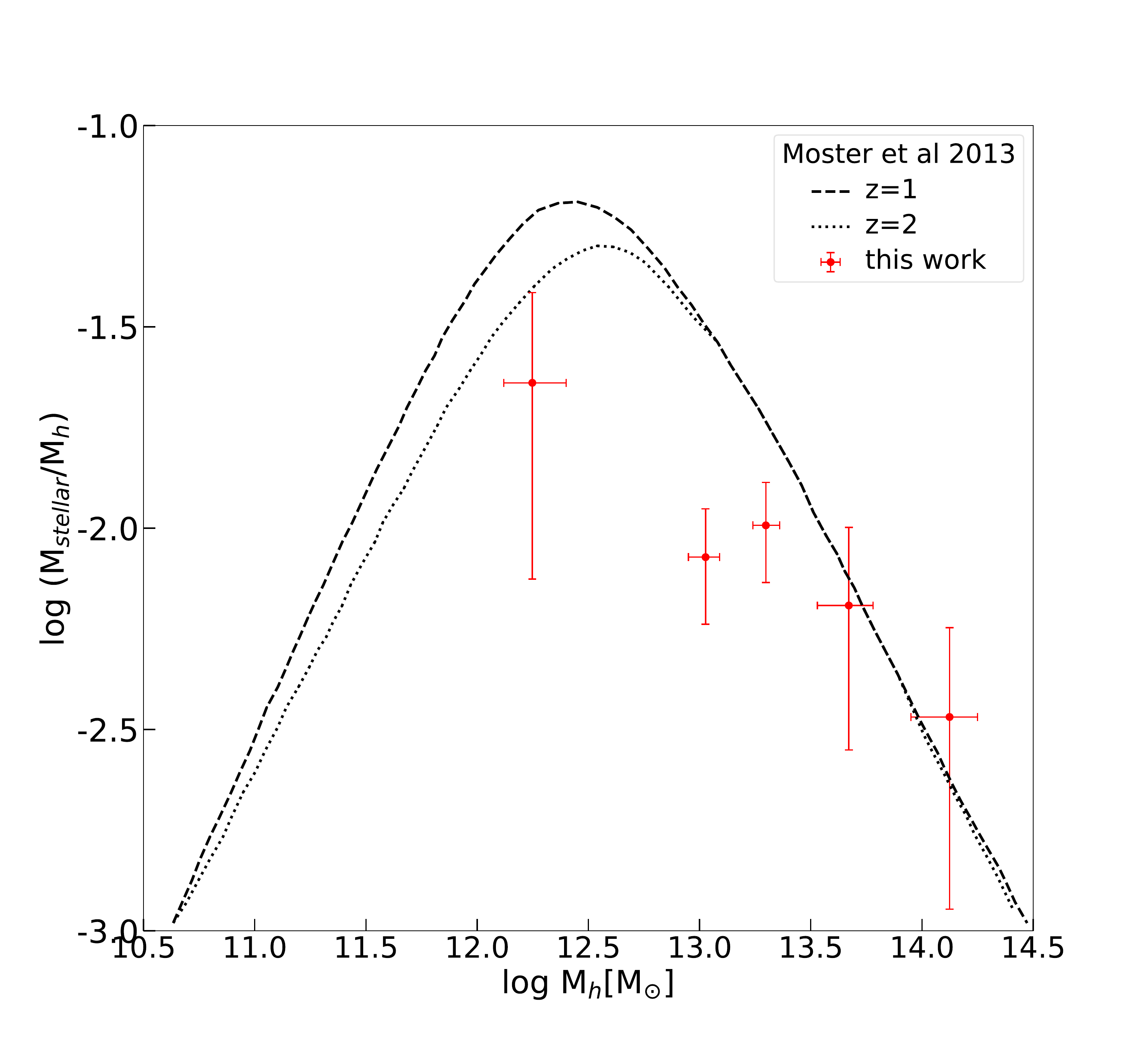}
\par\end{centering}
\protect\caption[Stellar mass-halo mass ratio (SHMR) for different PE $gzK_s$ sub-samples ]{\label{fig:Stellar-mass-halo-mass-model}Stellar mass-halo mass ratio (SHMR) for different stellar mass-selected  \PEgzK\ sub-samples at $z\sim1.6$ (filled red circles) as a function of halo mass. The measurements from our work are compared with model predictions by \citet{doi:10.1093/mnras/sts261} at $z=1$ and $z=2$, which are represented by dashed and dotted lines, respectively.}
\end{figure}

With the stellar masses and halo masses in hand, it is then straightforward to estimate the SHMR values for different \PEgzK\ subsamples, and we show these in Fig.~\ref{fig:Stellar-mass-halo-mass-model}.  As our UMPEGs are very massive, we are able to probe the high mass end of the stellar-to-halo mass relation. Our data show that the log(\Mstars$/M_{h}$) ranges from $\sim$$-2.5$ for the UMPEGs to $\sim$$-1.8$ for the less massive passive galaxies. It is clear that the SHMR decreases with increasing halo mass, indicating a reduced star formation efficiency in massive halos.  This trend has been seen before at at similar redshifts \citep[see][and references therein]{Legrand2019}, but most previous studies did not differentiate between star-forming and passive galaxies; moreover, our measurements robustly extend the SHMR to much larger stellar (and halo) masses than was previously probed at high redshift.

In Fig.~\ref{fig:Stellar-mass-halo-mass-model} we also compare our observed SHMR values with the results of numerical simulations by \citet[][black lines]{doi:10.1093/mnras/sts261}, which predict the SHMR for central galaxies of massive halos.  According to the  \citet{doi:10.1093/mnras/sts261} model, the SHMR reaches a peak at halo mass $\sim10^{12.5}\mbox{ }M_{\odot}$, while the lower SHMR values are due to different physical mechanisms that suppress star formation in the DM halo. Each process contributes differently at different mass. In the case of the low mass halos, feedback from supernova-driven winds \citep{doi:10.1093/mnras/169.2.229,1986ApJ...303...39D} is responsible for lowering the star forming efficiency. In contrast, processes such as feedback from active galactic nuclei (AGN; \citealt{2005Natur.435..629S,doi:10.1111/j.1365-2966.2006.10519.x,doi:10.1111/j.1365-2966.2005.09675.x}) and gravitational heating dominate in the massive halos.

The observed SHMRs for the more massive \PEgzK\ galaxies, including --- at the very massive end of our sample --- UPMEGs, agree well with the model predictions by \citet{doi:10.1093/mnras/sts261}.  In contrast, for the lower mass galaxies ($M_{200} \la 10^{13.5}M_{\odot}$), our measurements  are $\sim3-4$ times lower than the model predictions.  This discrepancy could be linked to an inefficient AGN and supernova feedback in quenched galaxies at intermediate masses. Alternatively, it is also possible that once quenched, intermediate-mass galaxies do not grow in stellar mass while their DM halos continue to grow, resulting in a lower SHMR than expected from models.  However, the most likely explanation is that the intermediate-mass \PEgzK\ galaxies may simply not be the central galaxies of their halos, but, rather, satellites in more massive environments. This last scenario is further supported by the detection of the one-halo term in the angular correlation measurements of the lower-mass \PEgzK\ galaxies at small separations \citep[see Sec.~\ref{sub:Error-Estimation_one_halo} and also][]{Sato2014}.  Meanwhile,  the agreement between the models and our observations at high masses gives further support to the idea that our UMPEGs are the central galaxies of their dark matter halos.

\subsection{Evolution of UMPEGs to $z\sim0$}

\begin{figure}
\begin{centering}
\includegraphics[width=8.5cm,keepaspectratio]{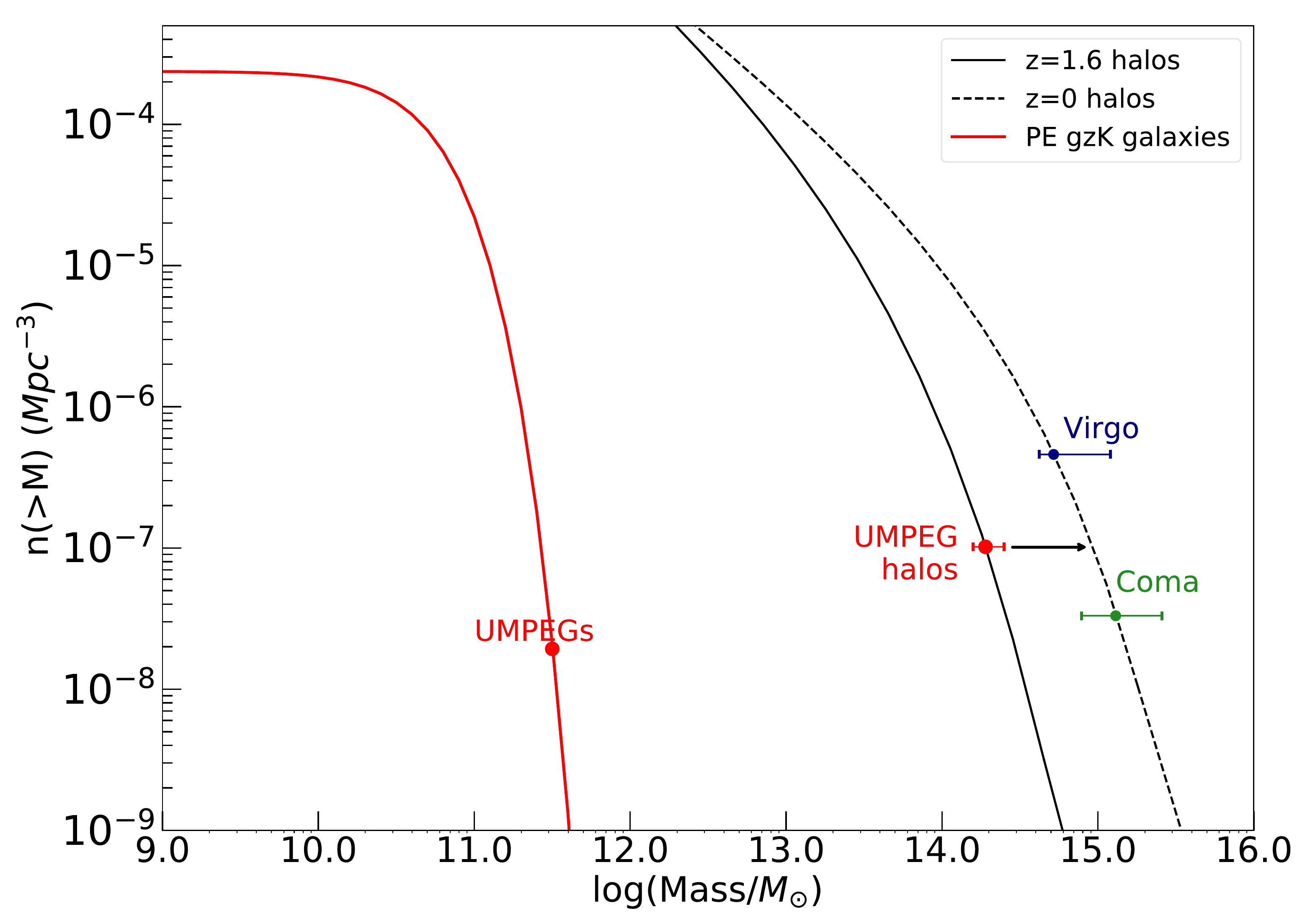}
\par\end{centering}
\protect\caption[Cumulative number density of the PE $gzK_s$ galaxies and DM halos for
the {\it MXXL} simulation.]{\label{fig:Cumulative_number_density}
Cumulative number densities of galaxies and halos as function of (stellar or halo) mass. The red curve shows the cumulative number density of \PEgzK\ galaxies from our study as a function of their stellar mass \citep{LARGE1}.   The black curves show the cumulative number densities of DM halos in the {\it MXXL} simulation at $z = 1.6$ (solid line) and $z=0$ (dashed).  The stellar masses of our UMPEGs are indicated with a point on the red curve, and their halo masses -- inferred from clustering -- on the black solid curve. The black arrow indicates the likely evolutionary path of the UMPEG halos from \zs1.6 to $z=0$ and assumes that halo rank order in mass is preserved over time. The present-day masses of the Virgo and Coma clusters are shown on the $z=0$ halo curve; for these clusters horizontal error bars represent the range of masses from different studies of Virgo \citep{1538-4357-512-1-L9,0067-0049-200-1-4,doi:10.1111/j.1365-2966.2011.18526.x}) and Coma \citep{GELLER897,0004-637X-671-2-1466,refIdgava}.}
\end{figure}

We next study the connection between observed galaxies and the simulated DM halos using a variation of the abundance matching technique \citep[e.g.,][]{0004-637X-717-1-379,0004-637X-696-1-620,doi:10.1111/j.1365-2966.2010.16341.x}.  Because we already have associated \PEgzK\  with their halos using clustering analysis, we can then apply abundance matching to ask how many massive halos contain massive quiescent galaxies. We focus our analysis here on the UMPEGs and use a simple approach in which the dark matter halos are assumed to grow in such a way that their rank-order does not change with time. We note that this is not ideal as the halo rank order may change over cosmic time due to, e.g., major halo-halo mergers. We also note that we focus on halo-halo clustering and ignore the clustering of sub-halos, an approach that could yield more detailed insights 
\citep[e.g.,][]{Vale2004, Conroy2006, ChavesMontero2016}, but which we cannot exploit here given our shallow dataset's inability to detect satellite galaxies around our UMPEGs. 

Indeed, as is found by M.~Sawicki et al.\  (submitted to MNRAS), UMPEGs are deficient in satellite galaxies with masses sufficient for detection in our Wide data.  This deficiency of satellites suggests major mergers as a growth mechanism in the UMPEGs' past (see M.~Sawicki et al.\ submitted to MNRAS), but makes subhalo clustering analyses impossible at present. Once deeper wide-field data are in hand, we intend to perform a more detailed analysis that uses halo merger trees to track the effect of rank-order reshuffling.  For now, howerver, given the limitations of our present dataset, we do not attempt to include sub-halos and satellites in our analysis and we restrict ourselves to the simple invariant halo rank-order abundance-matching approach.  Despite these limitations, our approach can nevertheless give us interesting insights into the nature of the \zs1.6 UMPEGs. 

As we discussed in Sec.~\ref{sec:haloMasses}, UMPEGs reside in some of the most massive halos at \zs1.6. In Fig.~\ref{fig:Cumulative_number_density} we plot the cumulative number density of halos from the \MXXL\ at \zs1.6 (solid black curve) and at \zs0 (dashed black curve); we mark with a red point the halo mass $M_h=1.6\times10^{14}M_\odot$,  which corresponds to an UMPEG with stellar mass \Mstars=10$^{11.4}$\Msun. We can then estimate the evolution of this ultra-massive halo to \zs0 by our simple abundance-matching argument: keeping number density constant at $n(>M) = 10^{-7}$Mpc$^{-3}$ between \zs1.6 and \zs0, we see that by \zs0 our \zs1.6 UMPEG halo grows to a mass of $M_h \sim 10^{15} M_\odot$ (black arrow in Fig.~\ref{fig:Cumulative_number_density}).  This \zs0 halo mass is comparable to the halo masses of local massive clusters of galaxies such as of Virgo \citep{1538-4357-512-1-L9,0067-0049-200-1-4,doi:10.1111/j.1365-2966.2011.18526.x} and Coma \citep{GELLER897,0004-637X-671-2-1466,refIdgava}, shown as blue and green points on the \zs0 halo mass curve in Fig.~\ref{fig:Cumulative_number_density}. 

Given the above argument, could UMPEGs be the direct progenitors of the most massive central galaxies of present-day massive clusters?   The stellar mass of the central galaxy in the Virgo cluster, NGC 4486,  is \Mstars=$10^{11.57}\mbox{ }M_{\odot}$ (\citealt{2013MNRAS.431.1405F}, using the \citealt{1538-3873-115-809-763} IMF). The stellar masses of the two central galaxies in the Coma cluster, NGC 4874 and NGC 4889, are $10^{11.98}\mbox{ }M_{\odot}$ and $10^{12.18}\mbox{ }M_{\odot}$ respectively \citep[][independent of IMF]{2017MNRAS.464..356V}.   Since UMPEGs already have stellar masses $>10^{11.4}\mbox{ }M_{\odot}$ at \zs1.6 \citep[i.e., 9.5 Gyr ago;][]{Wright2006}, it is plausible that they could become the massive central galaxies of low-$z$ massive clusters with only moderate growth via, e.g., minor mergers. Such growth scenario is compatible with simulations that predict that in the most massive halos much of the central galaxy stellar mass comes from satellite galaxies accreted at $z<2$ \citep[e.g.][]{doi:10.1093/mnras/sts261}.

We can investigate this question further by examining what fraction of the very massive halos associated with UMPEGs via our clustering analysis actually contains UMPEGs. For this we compare cumulative number densities, $n(>M)$ of ($M_\star=10^{11.4}M_\odot$) UMPEGs and of the corresponding halos ($M_h = 1.6\times10^{14}M_{\odot}$);  both these masses are marked with red points in Fig.~\ref{fig:Cumulative_number_density} on their corresponding cumulative mass functions.  The Figure shows that the UMPEGs have a comoving number density of $1.9\times10^{-8}$Mpc$^{-3}$, while the halos they are associated with  have a number density of $1.5\times10^{-7}$Mpc$^{-3}$.  There are therefore eight times more halos that are in principle capable of hosting UMPEGs than there are UMPEGs. This suggests that 7 in 8 of these most massive halos are likely to contain something other than an UMPEG:  either an ultra-massive {\it star-forming} galaxy or a group of lower-mass galaxies without a single UMPEG-mass central. In either of these two cases, it is then possible that UMPEGs are the descendants of these systems: in the former case an UMPEG could form by the quenching of star formation in the ultra-massive star-forming galaxy; in the latter case it could form through the merger of the lower-mass galaxies.  Here we note that $\sim$10\% of our UMPEGs are double-cored \citep[see][]{LARGE1};  these double-cored systems could represent recent or ongoing mergers and would support the idea of UMPEG formation through the merger of lower-mass galaxies. Further evidence for a major-merger UMPEG formation scenario, albeit with the mergers happening at earlier times, $z>1.6$, comes from the mass gap seen between UMPEGs and their most massive satellites (see M.~Sawicki et al., submitted to MNRAS). 

In the context of dusty massive starbursts, it is important to note that our UMPEGs cluster much more strongly than do sub-millimetre galaxies \citep[SMGs, e.g.,][]{Blain2004, Hickox2012, Wilkinson2017} and Dust Obscured Galaxies \citep[DOGs, e.g.,][]{Brodwin2008, Toba2017}.  Consequently, it seems that UMPEGs are associated with more massive, rarer dark matter halos than those that host typical high-$z$ starbursts.  This fact does not rule out the possibility that some dusty starbursts are the direct progenitors of UMPEGs, of course, but it does suggest that typical SMGs and DOGs inhabit lower-mass structures than the UMPEGs and that most of them will not become UMPEGs at later times simply by quenching.

In summary, the picture that is emerging from our analysis is that UMPEGs may be the direct progenitors of some ($\sim$ 1 in 8) of the central galaxies of present-day massive clusters.  This is because they appear to have very high stellar masses while their very strong clustering resembles the clustering strengths of (proto)clusters at similar and higher redshifts \citep{Papovich2008, Rettura2014, Toshikawa2018}.  It is less clear from our clustering analysis alone what are the direct progenitors of the UMPEGs, although the weaker clustering strengths of high-$z$ dusty starbursts suggest that most of those objects do not evolve to become UMPEGs.  Alternatively UMPEGs could have formed via the merging of lower-mass galaxies already present in distant proto-clusters \citep[e.g.,][]{Ouchi2005, Lemaux2009, Toshikawa2012, Jiang2018, Oteo2018} -- a scenario we explore further by studying the proximate environments of our UMPEGs in M.~Sawicki et al.\ (MNRAS, submitted).  Of note is that only some ($\sim$1 in 8) of the present-day cluster central galaxies were already very massive and quiescent at \zs1.6, while $\sim$7 of 8 protocluster-mass halos must still contain either an ultra-massive star-forming progenitor or a set of building-block components still destined to merge.

\section{Conclusions}\label{sec:summary}

Using a sample of  massive quiescent \PEgzK\ galaxies at $z\sim1.6$ drawn from a dataset of unprecedented area, we used clustering measurements to link the properties of the galaxies, split into subsamples by their \Ks\ magnitude, to those of their dark matter halos.  The subsamples range from \Ks\ $\sim$23 to 19.5, corresponding to stellar masses from \Mstars$\sim 10^{10.3}$\Msun\ to $\sim 10^{11.5}$\Msun.  The brightest, most massive subsample -- the UMPEGs -- form the special focus of this work. 

We presented the two-point angular correlation functions for the passive galaxy subsamples, together with the best power-law fits. Using the observed redshift distributions of these galaxies, we de-projected the spatial correlation functions from the angular ones, and estimated correlation lengths for the UMPEGs as well as for the lower-mass \PEgzK\ galaxies. By comparing our clustering measurements to those of the DM halos from the Millennium XXL simulation, we then estimated the halo masses for the \PEgzK\ galaxy host halos, including those of the UMPEG-hosting halos,  as a function of galaxy stellar mass.  

Our primary results are as follows:

\begin{enumerate}

\item We derived the correlation length, $r_{0}$, for the UMPEGs and found that the UMPEGs have very strong clustering, stronger than that for any other galaxy population at high redshift and comparable to that of massive high-$z$ (proto)clusters.

\item We also confirmed previous findings that the correlation length for the clustering of lower-mass \PEgzK\ galaxies is dependent on their $K_{s}$ magnitude. In addition to this luminosity dependence, there is a clear enhancement in the clustering of the (lower-mass) passive galaxies at small scales, as also found by \citet{Sato2014}. This  ``one-halo term'' enhancement is suggestive of multiple quiescent  \PEgzK\ galaxies residing in the same dark matter halo. 

\item Comparing our simple clustering observations with the clustering measurements of DM halos from the Millennium XXL simulation \citep{doi:10.1111/j.1365-2966.2012.21830.x}, we constrained UMPEG halo masses and concluded that the UMPEGs inhabit some of the most massive ($M_h \sim10^{14.1}h^{-1}M_{\odot}$) dark matter halos at $z \sim 1.6$.

\item  Given that DM halos grow over time, UMPEG halos are likely to grow. Our simple halo abundance matching analysis, which assumes that the halo mass rank order is preserved over time, suggests that by $z \sim 0$ UMPEG halos may grow to a typical mass of $M_h \sim10^{15}M_{\odot}$. This halo mass is comparable to that of massive $z\sim 0$ clusters such as Virgo and Coma. The descendants of UMPEGs may thus reside in massive galaxy clusters today, and given their large $z \sim 1.6$ masses, may be the progenitors of some ($\sim$ 1 in 8 from abundance arguments) of the massive cluster central galaxies at $z\sim 0$. 

\item We studied the SHMR of our massive passive galaxies. Our measurements for the massive (UMPEG) end of the mass distribution are in good agreement with the SHMR models of \citet{doi:10.1093/mnras/sts261}. However, there is a discrepancy with the models at lower masses that could be caused by inefficient feedback in the models as compared to \PEgzK\ galaxies, by a divergence of the halo and galaxy growth rates after the quenching of star formation, or -- most likely, we feel -- due to multiple galaxies (passive or star-forming) present within the same halo.

\end{enumerate}

Overall, based on their very strong clustering, we conclude that the most massive passive galaxies (UMPEGs, \Mstars$>10^{11.4}$\Msun) at \zs1.6 are likely to be the central galaxies of some ($\sim$ 1 in 8) of the massive ($\sim$$10^{14.1}h^{-1}M_{\odot}$) high-$z$ protoclusters. They are likely to evolve into some of the massive central galaxies of present-day $\sim10^{15}M_{\odot}$  massive clusters, although most ($\sim$7 out of 8) present-day massive cluster progenitors do not yet have such an ultra-massive quiescent central galaxy at \zs1.6.

\section*{Acknowledgments}

We thank Ivana Damjanov, Bobby Sorba, Rob Thacker, and the anonymous referee for useful suggestions, and the Natural Sciences and Engineering Research Council (NSERC) of Canada for financial support.

This  work  is  based  on  observations  obtained  withMegaPrime/MegaCam, a joint project of CFHT and CEA/DAPNIA, at the Canada-France-Hawaii Telescope (CFHT) which is operatedby the National Research Council (NRC) of Canada, the Institut National des Science de l'Univers of the Centre National de la Recherche Scientifique (CNRS) of France, and the University of Hawaii. This work uses data products from TERAPIX and theCanadian Astronomy Data Centre. It makes use of the VIPERS-MLS database, operated at CeSAM/LAM, Marseille, France. This work is based in part on observations obtained with WIRCam, a joint project of CFHT, Taiwan, Korea, Canada and France. The research was carried out using computing resources from ACEnet and Compute Canada.





\bibliographystyle{mnras}
\bibliography{gzKclustering} 












\bsp	
\label{lastpage}
\end{document}